\documentclass[fleqn,usenatbib]{mnras}

\usepackage{newtxtext,newtxmath}

\usepackage[T1]{fontenc}
\usepackage{ae,aecompl}

\usepackage{graphicx}	
\usepackage{amsmath}	
\usepackage{amssymb}	
\usepackage{nccmath}
\usepackage{subcaption}
\usepackage{xcolor}
\usepackage{multirow}
\usepackage{floatrow}
\usepackage{caption}
\usepackage{placeins}
\usepackage{float}
\usepackage{afterpage}
\usepackage{enumerate}
\title[Pattern recognition with SA]{A Simulated Annealing algorithm to quantify patterns in astronomical data}

\author[M. Chira \& M. Plionis]
{Maria Chira$^{1,2}$,\thanks{E-mail: mchira@physics.auth.gr}
Manolis Plionis$^{2,1}$
\\
$^{1}$Physics Dept., Aristotle Univ. of Thessaloniki, Thessaloniki, 54124, Greece\\
$^{2}$National Observatory of Athens, P.Pendeli, Athens, Greece\\
}

\date{Accepted 2019 October 9. Received 2019 October 9; in original form 2019 March 21}

\pubyear{2019}

\begin{document}
\label{firstpage}
\pagerange{\pageref{firstpage}--\pageref{lastpage}}
\maketitle

\begin{abstract}
We develop an optimization algorithm, using simulated annealing for the  quantification of patterns in astronomical data based on techniques developed for robotic vision applications. The methodology falls in the category of {\em cost minimization algorithms} and it is based on user-determined interaction - among the pattern elements - criteria which define the properties of the sought structures. We applied the algorithm on a large variety of mock images and we constrained the free parameters; $\alpha$ and $k$, which express the amount of noise in the image and how strictly the algorithm seeks for cocircular structures, respectively. We find that the two parameters are interrelated and also that, independently of the pattern properties, an appropriate selection for most of the images would be $\log{k} =-2$ and $0<\alpha \lesssim 0.04$. The width of the effective $\alpha$-range, for different values of $k$, is reduced when more interaction coefficients are taken into account for the definition of the patterns of interest.  Finally, we applied the algorithm on {\em N}-body simulation dark-matter halo data and on the {\em HST} image of the lensing {\em Abell 2218} cluster to conclude that this versatile technique could be applied for the quantification of structure and for identifying coherence in astronomical patterns. 
\end{abstract}




\section{Introduction}

\label{S:1}

Simulated annealing (SA) is an optimization algorithm introduced by \citet{Kirk1983}. The idea behind this technique is based on the annealing of solids, which is a physical process leading to the crystallization of solid materials. During the annealing, the solid material is heated up to a melting temperature so that the atomic energy increases and atoms move freely. While the material is gradually and slowly cooled, the atoms at each temperature tend to rearrange so that the energy of the configuration decreases. At each temperature the material should reach thermal equilibrium and the procedure is repeated until the material reaches a "freezing" temperature, which corresponds to the state of minimum energy. If the cooling of the material is rapid (rapid quenching) then the above is not attained and the process results in a higher energy, polycrystalline atomic structure.

Combinatorial optimization approaches are used in many research fields such as medicine \citep[c.f.][]{Keelan2018}, engineering \citep[c.f.][]{Sonmez2007}, AI \citep[c.f.][]{Eckrot2017}, image processing  \citep[c.f.][]{Storvik1994, STOICA2005}, physics and astronomy \citep[c.f.][]{Habib2006,Stoica2008,Tempel2014,Kovalenko2017,Tempel2018}. In such applications the optimum solution corresponds to that of the minimum energy state yearned during the annealing of solids, while the cost function, corresponding to the quantity to be minimized (e.g., cost of a construction, weight or volume of structures, time required for completing a task or any energy-like parameter), is analogous to the energy of the material. The methodology of SA requires the definition of a cost function depending on the nature of the problem, a cooling schedule of the temperature (or a quantity analogous to the temperature) which will ensure that the algorithm will escape local minima and result to the global minimum of the cost function, i.e. the optimum solution. 

In this work, we study the details of the SA methodology, based on the approach of \citet{HH1993} used in robotic vision, as a tool in order to develop a specific pattern-identification algorithm for various astronomical problems. Specifically, the particular approach is focused on the detection of structure, such as circles, arcs, lines, in images with background noise. Such structures are commonly found in nature and their detection in images can lead to the identification of physical phenomena in a variety of scales. Especially in astronomy, various phenomena and processes are connected with the presence of roughly circular or curvy structures. From larger to smaller scales we refer to examples where automatic detection of physically connected structures is useful and could probably lead to an easier statistical quantification of connected physical phenomena in large volumes of data. Typical examples of such phenomena are as follows:

\begin{itemize}
\item {\em The cosmic web:}
The distribution of matter in the Universe is far from uniform. Instead, dark matter (DM) halos, the cosmic building blocks, and consequently galaxies form what is known as the {\lq cosmic web\rq} \citep{Bond1996}. This hierarchical cosmic structure  consists of four main web-elements, clusters, filaments, sheets and voids. Matter flows through filaments into clusters which are formed at their intersections. A large number of filament-finding algorithms have been developed and applied to 2D and 3D galaxy catalogues and N-body simulations through the years \citep[for a review, see][]{Libeskind2018}. Among the different algorithms, some have also incorporated a simulated-annealing {\lq engine\rq}, as the one based in the Bisous model of \citet{STOICA2005}. This methodology has been applied on galaxy catalogues from redshift surveys like the 2dF \citep{Stoica2008}, the SDSS \citep{Tempel2014, Tempel2016} and also on simulated data \citep[e.g.,][]{Tempel2014b, Veena2019}. However, it is well established that using different approaches for cosmic-structure detection leads to identification of structure with different statistical properties and also to different mass and/or volume fraction of the web-element types in the cosmic web \citep[for a comparative study on the results of different web-element-identification algorithms, including the Bisous method, see the review of][]{Libeskind2018}. Therefore, there is ample space for further development of cosmic structure finders in order to understand the pros and cons of the different methodologies.\\

\item{\em Gravitational lensing:}
 According to General Relativity, the curvature of spacetime affects the direction of light propagation. A consequence of the bending of light, in the presence of curvature, is the phenomenon of gravitational lensing, which occurs when the electromagnetic waves emitted by a background source encounter the strong gravitational potential of a massive object (e.g. galaxy or cluster). The latter plays the role of the lens and, depending on the relative position of the two objects, the bended light of the background source forms a ring, known as {\lq Einstein's ring\rq}, or separate, cocircular or roughly so, arcs. In most cases, the lensed light has such geometrical characteristics that could be detected by the SA pattern recognition algorithm \citep[for further information on gravitational lensing, see reviews of][]{Refsdal1994,Bartelmann2010}.\\
 
\item{\em Filaments in molecular clouds:}
 During the last decades, the presence of filaments in molecular clouds was predicted by numerical simulations and it is nowadays well established via observations \citep[for a review on molecular clouds, see][]{Andre2013}. The filamentary structure with common properties found in every interstellar cloud has caused the attention of researchers in the field of stellar and planetary formation, since the link between the structure of the ISM and star formation is considered a crucial key point for understanding the relevant physical properties in early stages of formation. As in both previous cases, although the physics and the scale of the problem is different, the method discussed in this work could be extended in order to be applied on images of molecular clouds to quantify their filamentariness.\\
 
 \end{itemize}
 
At this point, we need to underline that SA is only one of the several existing optimization techniques among which, e.g., Genetic Algorithms, Tabu Search, Neural Networks, Ant Colony and Particle Swarm Optimization (PSO) \citep{Keelan2018,Pham}. A major advantage of SA is its simplicity and versatility. The optimization procedure is able to find the optimum solution of a vast variety of problems, requiring only the appropriate definition of the cost function and selection of cooling schedule. Different criteria and conditions, depending on each specific problem of interest, are simply represented by extra terms of the cost function leaving the optimization procedure intact.
 
 A particular outcome of different physical processes, that shape small- and large-scale cosmic structures as well as other astronomical configurations, which we wish to exploit in attempting to reveal patterns, is the different type of statistical alignments. One such case is that of DM halos and/or galaxies and clusters within large-scale structures, them being filaments or sheets \citep[cf.][]{Binggeli1982,Plionis1994,Faltenbacher2002,Kasun2005,Joachimi2015,Pandya2019}. Other type of alignments are also expected due to gravitational lensing \citep[e.g.,][]{Bartelmann2010}.


\section{Method}
\label{S:2}

We base our methodology for pattern recognition on the techniques proposed by \citet{HH1993}, based in robotic vision. We build a cost-minimization algorithm, considering the image elements as being particles of an interacting spin system. The physical quantity to be minimized, i.e. the cost function, is the energy of the system as it is calculated for a system of $N$ interacting spins taking values $\sigma_i \in \{-1,1\}$. A configuration of the system is characterised by a vector containing the $N$ states of each spin $\vec{\sigma} =  [\sigma_1,...\sigma_N]$. 

The total energy of the system of N interacting spins, as a function of $\sigma$'s \citep[for the detailed calculations see][]{HH1993}, can be written as:  

\begin{ceqn}
\begin{align}
E_(\vec{\sigma}) = \frac{1}{4}\LARGE( -\sum_{i=1}^{N}\sum_{j=1}^{N}c_{ij} + \alpha N^2 \LARGE) -  \sum_{i=1}^{N}\sum_{j=1}^{N}\frac{1}{2}(c_{ij}-\alpha)\sigma_i \sigma_j \notag \\
- \sum_{i=1}^{N}\frac{1}{2}\LARGE( -N \alpha + \sum_{j=1}^{N}c_{ij}\LARGE) \sigma_i \;\;,
 \label{energy}
\end{align}
\end{ceqn}
where $\alpha$ is a positive parameter which is related to the percentage of noise elements in the image, i.e. to the signal to noise (SN) ratio and $c_{ij}$ denotes the interaction coefficient between elements $i$ and $j$. The choice of the interaction coefficients depends on the physical and geometrical properties of the system under study and they can differ depending on the details of the patterns that one wishes to identify.

The goal of the algorithm is to assign a value, $-1$ or $1$, to every spin $\sigma_i$, with $\sigma = 1$ characterizing the elements that are part of the seeked pattern and $\sigma = -1$ the noise elements, so that the cost function of equation (\ref{energy}) is minimized. The minimization procedure and the way of expressing the spin interactions via the interaction coefficients are described in the following paragraphs.

\subsection{Calculation of spin-interactions: Individual interaction coefficients.} \label{inv_params}

The main goal of applying the above described method is the detection of patterns in real and simulated images, which would correspond to physically related structures based on specific {\lq interaction\rq} properties. Interactions between the image elements are expressed via the spin interaction coefficient, $c_{ij}$, $i,j = 1,...,N$, which is a positive number taking values $\in [0,1]$, with $c_{i,j}\rightarrow 0$ corresponding to no interaction, while higher values of $c_{ij}$, ($c_{i,j}\rightarrow 1$), denote higher interactions between spins $i$ and $j$, i.e. high {\em co-shapeness}.

 Each spin (or image element) is characterized by a variety of physical properties, among which the most important are its position and its direction, i.e. by the Cartesian coordinates (in 2D or 3D) and the position angle formed by the image element and the horizontal axis, $xx'$, of the Cartesian coordinates system. Important criteria for the identification of patterns must depend on the underlying physical interactions of the elements constituting the pattern. Usual criteria include the spatial separation of the image elements, their geometrical properties or physical properties like colour, weight, individual element shape, topological properties etc.
\subsubsection{Physical criteria for the characterization of an interacting structure.}

The astronomical applications of the proposed methodology, as explained in the introduction, are related to the detection of geometrical patterns like curves, circles, filaments, etc., and thus we primarily consider that two spins are interacting if they are close and cocircular. These two factors are quantified via the {\em proximity coefficient} and the {\em cocircularity coefficient}.  Moreover, as we explain in detail in the next paragraphs, we attempt to improve the efficiency of our method by taking into account more criteria, such as the {\em smoothness} of the curve defined by the two spins via the {\em smoothness coefficient} and the {\em mass coefficient} depending on the masses of the elements. The total interaction coefficient is equal to the product of the individual coefficients. We now present the list of the basic coefficients used for the current application of the method; most already discussed in \citet{HH1993}.

\begin{figure}
\includegraphics[width=\hsize]{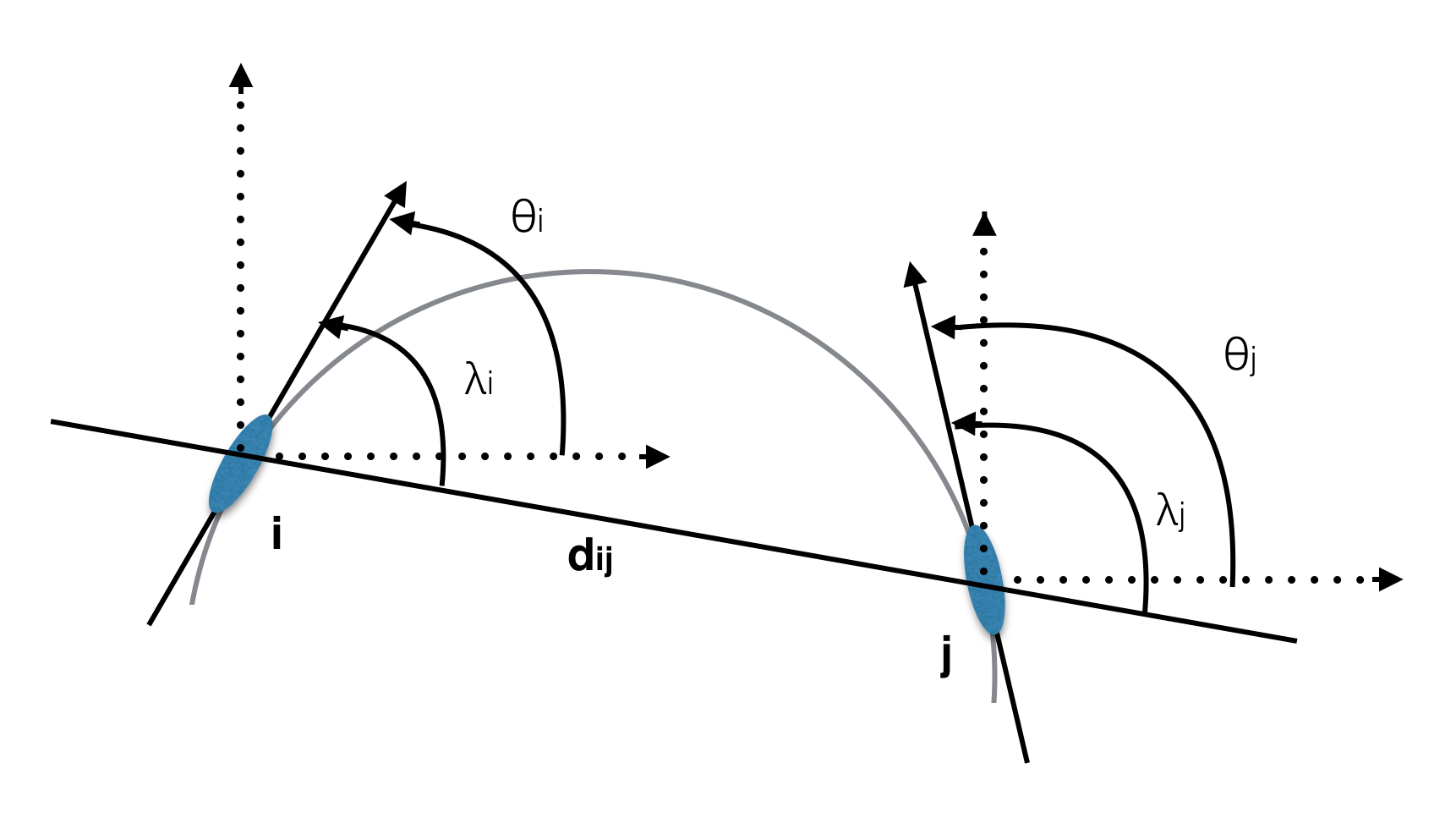}\caption{Two image elements $i,j$ spanning a distance $d_{ij}$. Their directions are shown by the arrows [figure following \citet[fig.3]{HH1993}].}\label{cocir}
\end{figure}

\subsubsection*{Proximity coefficient}

The {\em proximity coefficient}, which weights the distance between two elements, is simply defined as

\begin{ceqn}
\begin{align}
  c_{ij}^{prox} = \exp(-d_{ij}^{2}/ 2\sigma_{d}^{2}),
   \label{prox}
\end{align}
\end{ceqn}
with $d_{ij}$ being the spatial separation between two image elements, $\sigma_{d}$ being the standard deviation of separations in the image. The exponential relation is used as more effective in weighting the near-neighbor elements. Note that $c_{ij}^{prox} \in [0,1]$.

\subsubsection*{Cocircularity coefficient}

The {\em cocircularity} coefficient is defined as 

\begin{ceqn}
\begin{align}
  c_{ij}^{cocir} = (1 - \Delta_{ij}^{2}/\pi^{2})\exp(-\Delta_{ij}^{2}/ k),
   \label{cocirc}
\end{align}
\end{ceqn}
where $\lambda_i$, $\lambda_j$ are the angles formed by the directions of the image elements $i,j$, respectively, and the straight line connecting the image elements and  $\Delta_{ij} = |\lambda_i + \lambda_j - \pi|$. In Fig. \ref{cocir} the image elements $i$ and $j$ have directions denoted by the arrows in the image. Note however that for our purposes, antiparallel tangent vectors are equivalent, i.e. it is the line defined by the tangent that is important in our applications. For example, in an image of galaxies or haloes, the tangent would be defined by the major axis of the 2D-fitted projected ellipsoid, which is a line segment for which direction is not meaningful. Thus, the direction of the tangent-vector chosen is such that the $\lambda$ angles, as these are defined anticlockwise from the line connecting the two image elements to each tangent, have $\lambda \in [0 \deg,180 \deg)$. Here again $c_{ij}^{cocir} \in [0,1]$ with $1$ corresponding to the image elements being part of a circle, i.e. being tangent to the same circle. 

\subsubsection*{Smoothness coefficient}\label{smooth_definition}

\begin{figure}
\includegraphics[width=\hsize]{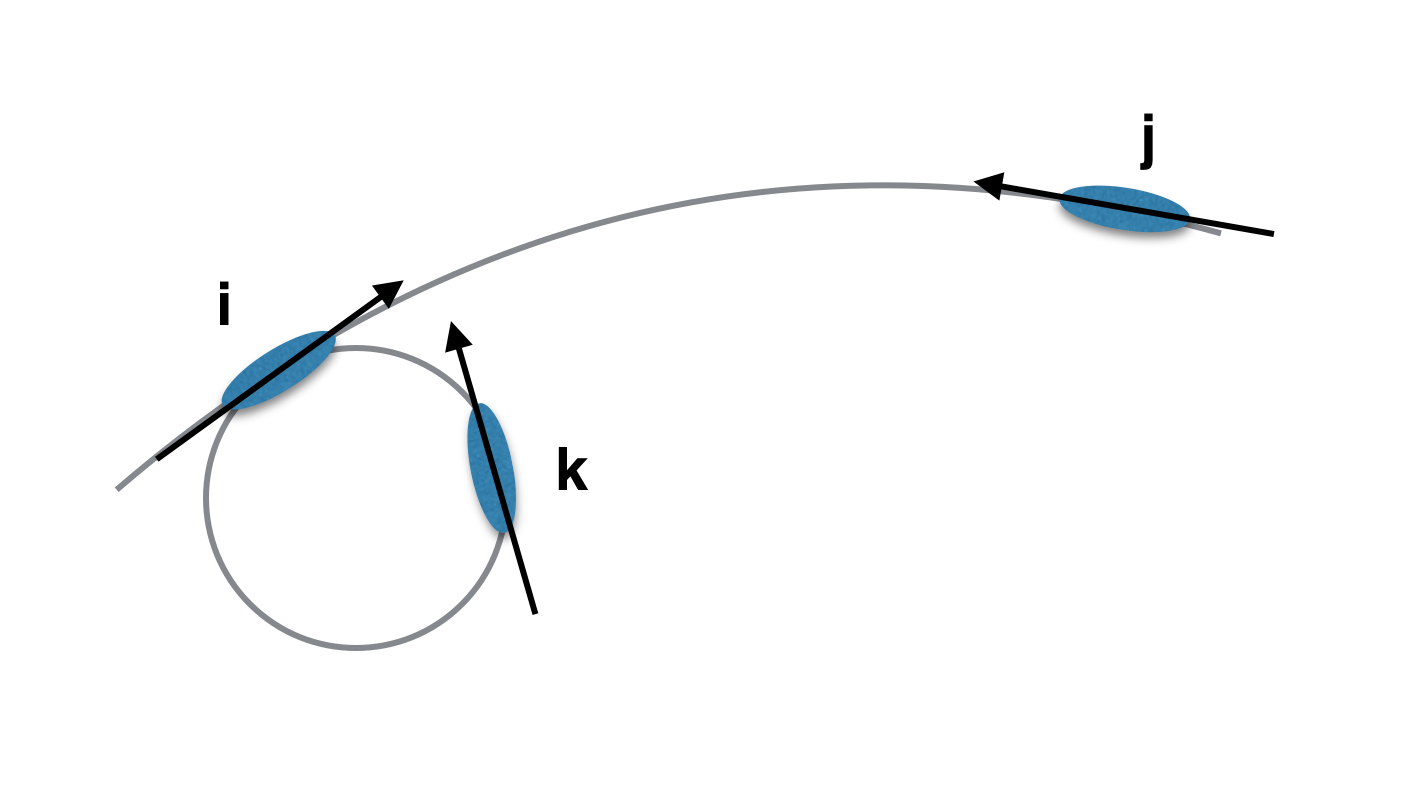}
\caption{Explanation of the {\em smoothness coefficient} [the figure following \citet[fig.4]{HH1993}].}\label{smooth}
\end{figure}
The role of the {\em smoothness coefficient} is to favor smooth curves and disfavor sharp ones and also to avoid connecting image elements belonging to parallel curves (see the right-hand panel of Fig. \ref{smooth}). It is defined as

\begin{ceqn}
\begin{align}
  c_{ij}^{smooth} = (1 - \lambda_i(\pi - \lambda_i)/\pi^{2})(1 - \lambda_j(\pi - \lambda_j)/\pi^{2})
   \label{sm_coef}
\end{align}
\end{ceqn}
As an example in Fig. \ref{smooth} (left-hand panel), for the case of spins $i,j$ and $k$, the {\em smoothness coefficient} would take a larger value for $i,j$ than for $i,k$.

\subsubsection*{Mass coefficient}\label{mass_definition}

One other possible criterion for interaction, which can be quite meaningful in many physical problems, would be the mass of the pair elements. The {\em mass coefficient} (which is discussed in detail in the next paragraphs) can be defined as:

\begin{ceqn}
\begin{align}
c_{i,j}^{mass} = \frac{M_{i}M_{j}}{M_{max}^{2}},
 \label{ms}
\end{align}
\end{ceqn}
where $M_{i}, M_{j}$ are the masses of the members of the pair $({i,j})$ and $M_{max}$ is the mass of the most massive element in the image, used to normalize the coefficient to the range $[0,1]$.

In a similar manner, depending on the nature of the image, the relevant physics, and the patterns of interest, more coefficients could be defined in order to insert more or alternative interaction criteria for the definition of a pattern.

\subsection{Minimization procedure: SA}\label{SA}

The SA algorithm applied here is based on the Monte Carlo algorithm described by Metropolis et al. (1953). The initial configuration of the spin system at an initial temperature, $T$, is $\vec{\sigma_0} = [\sigma_1,...\sigma_{i0}...\sigma_N]$, where $\sigma_{i= 1,...,N}$ are randomly assigned to either values, $1$ or $-1$. The algorithm in each step proposes a transition for a randomly selected spin, $i_0 \in \{1,...,N\}$, and this transition is accepted with a probability:

\begin{equation}
  P=\begin{cases}
    1, & \text{if $\Delta E<0$}.\\
    \exp(-\Delta E /T), & \text{otherwise.}
  \end{cases}
\end{equation}
$\Delta E$ is the change of the energy of the system imposed by the system transition, when a spin is flipped from $\sigma_{i0}\rightarrow - \sigma_{i0}$   which is calculated with the use of equation \ref{energy} as (for a detailed calculation of the change of energy see \cite{HH1993}):

\begin{ceqn}
\begin{align}
  \Delta E _{\sigma_{i_0}\rightarrow - \sigma_{i_0}} = 2\sigma_{i_0} \left[\sum_{j=1,j\neq i_0}^{N}(\lambda-c_{i_0j})\sigma_j) + (N\lambda - \sum_{j=1}^{N} c_{i_0j})\right].
   \label{de}
\end{align}
\end{ceqn}

The procedure of proposing transitions is repeated $100N$ times or until $10N$ transitions are accepted when, for the Metropolis procedure, the system is considered to have reached a near equilibrium state at this particular initial temperature, $T$. As it is pointed out in \citet{Kovalenko2017}, it  has been proved by \citet{Geman1984} and \citet{STOICA2005} that appropriate logarithmic cooling schedules can ensure the convergence to a global minimum. However, since establishing such a schedule for probabilistic problems is not a trivial task, in our approach we adopt the cooling schedule of \citet{HH1993} which is also similar to that selected by \citet{Kovalenko2017}. Following this schedule, the temperature of the system is lowered according to $T_{new} = 0.93 T_{old}$ and then the previous steps are repeated.

The algorithm stops when the system is {\lq frozen\rq}, which occurs when less than $1\%$ of the proposed transitions are accepted. The final configuration of the system, at this stage, characterizes the image elements either as {\lq pattern elements\rq} (those with $\sigma_i = 1$)
or as {\lq noise elements\rq} (those with $\sigma_i = -1$).

\section{Application of the Method on Mock images}\label{Application_Mock}
\subsection{Data used and test-bed images}
\begin{figure}
   \centering
   \includegraphics[width=0.98\textwidth]{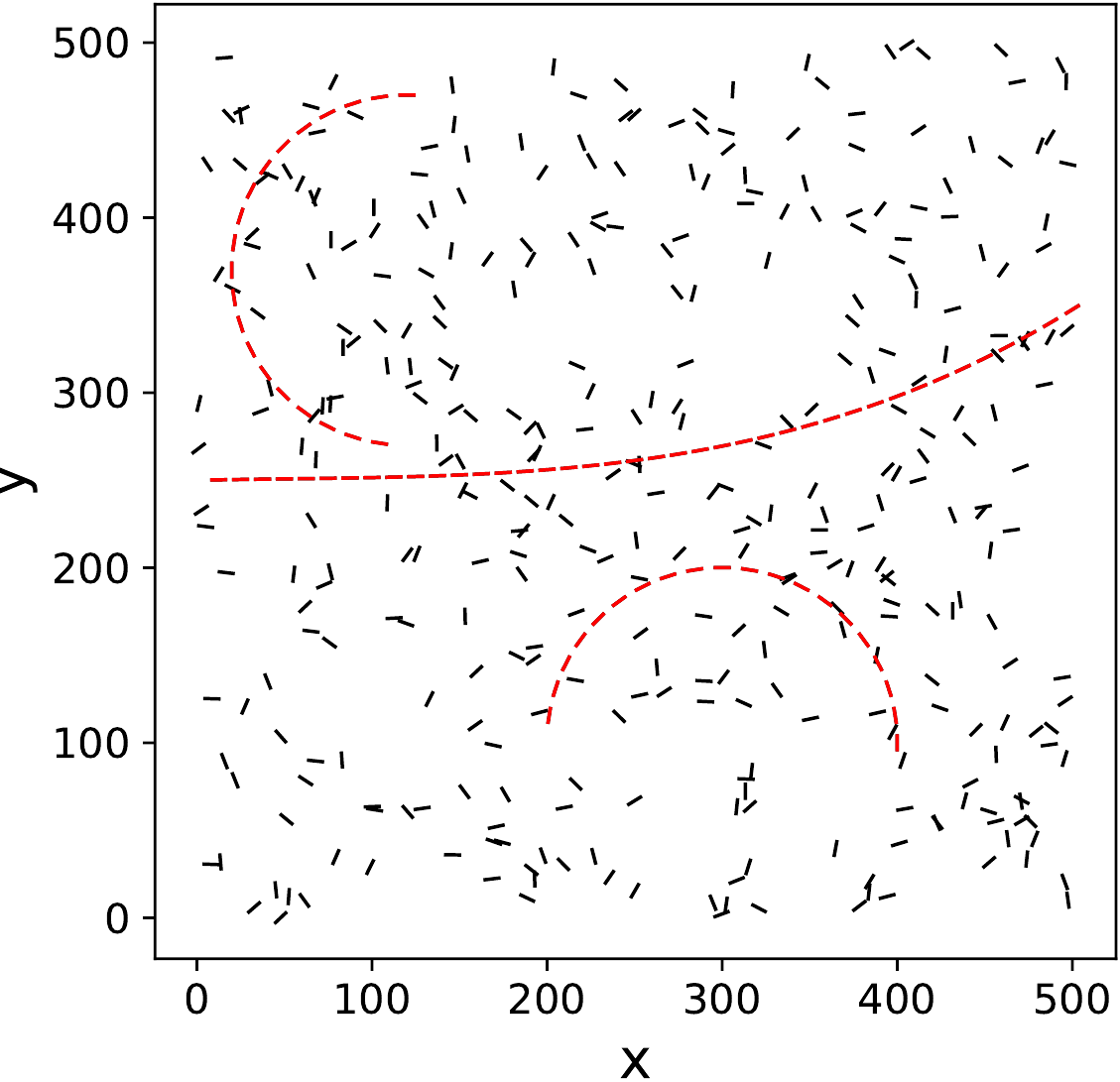}
      \caption{The main test-bed figure with 440 elements: 80 elements constituting the pattern, in red, and 360 noise elements, in black. We tag this test-bed as $\rm P1$.            }
         \label{pat440}
   \end{figure}

In order to apply and test this method, we produce a number of mock images containing curvy patterns. For constructing a realistic image background, we use the Monte Carlo method in order to reproduce the abundance function (AF) of DM halos, extracted from the $\Lambda$ CDM halo catalog of the light-cone N-body simulations of the {\lq dark
energy universe simulation\rq } (DEUS) project \citep{Alimi2010,Rasera2010,Courtin2011} in \cite{Chira2018}. The number density of the background is approximately that of the 2D projection of a $50$ Mpc h$^{-1}$-thick N-body simulation slice when the mass limit is set to $10^{14} M_{\odot}$, which is $d_{2D}\approx 1.8\times 10^{-3}$. The set of mock images that we work on consist of the background and of a variety of circular patterns of different levels of complexity. We start by applying the algorithm on simple cases, i.e. images containing a circle or a curve with a high density of image elements and we increase the complexity by putting different patterns in the same image. An example is shown in Fig.\ref{pat440}, where the background consists of 360 noise image elements and 80 pattern elements which define two hemispheres and a curve. We will eventually also apply the algorithm on projected images of the cluster {\em Abell 1656} consisting of 419 SDSS galaxies \citep{Ahn2014} with r-magnitude $<17.77$ and mock circular patterns constructed in an attempt to simulate strong lensing images.

\subsection{Success evaluation procedure}
   
By using a number of mock images for which we know a priori which of the elements are part of a pattern, we can evaluate the success of the algorithm in identifying the pattern. As main criteria for such an evaluation, we set
\begin{itemize}
\item[(a)]the number of the {\lq false\rq} elements, i.e., the number of elements which the algorithm falsely tags as pattern-elements, and 
\item[(b)]the number of {\lq missed\rq} elements, i.e., the number of pattern elements which the algorithm falsely identifies as noise elements. \end{itemize}
This procedure can be used in order to assess the effective range of values of the {\lq free\rq} parameters of the model interactions, i.e. parameters $\alpha,k$.

\subsubsection{Parameter $\alpha$} 
\begin{figure}
\includegraphics[width=1.1\columnwidth]{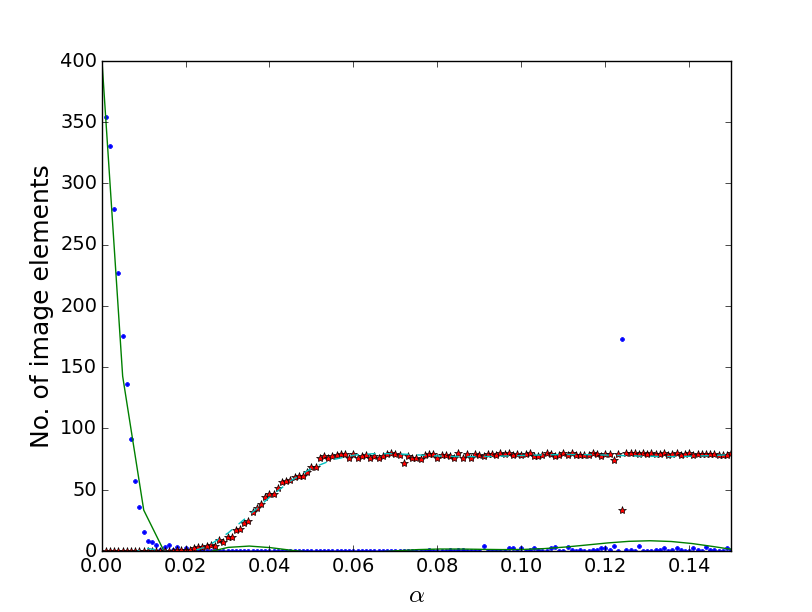}\vfill
\caption{The number of {\lq false\rq} elements (blue dots) and {\lq missed\rq} elements (red stars) and the fitted polynomials (green solid and blue dashed line respectively). The polynomials are fitted locally in the range of values approximately $\pm 0.02$ around the intercept point of the two curves to find the optimum $\alpha$-value.
}
\label{fit}
\end{figure}

The first important issue is to clarify the effect  of the $\alpha$ parameter on the results and which criteria can define its optimum value, to be used for eventually applying the algorithm on images with unknown properties. Thus, we run a first test using the test-bed of Fig. \ref{pat440} to investigate how the number of {\lq false\rq} (blue) and {\lq missed\rq} (red) elements (defined previously) is affected by the value of the $\alpha$ parameter. A typical behavior of these quantities is presented in Fig. \ref{fit} where, for small values of $\alpha$, the number of {\lq false\rq} elements is large, while no pattern elements are missed. For values of the order of $\alpha\in[0.015,0.025]$, the number of {\lq false\rq} elements drops dramatically while for higher $\alpha$-values the algorithm starts to miss pattern elements. As the value of $\alpha$ increases even more ($\alpha \gtrsim 0.05$) the algorithm misses all the pattern-elements. Obviously, the ideal case would be a value of $\alpha$ providing $0$ {\lq missed\rq} and $0$ {\lq false\rq} elements. Such a value can be found for many of our mock images used. However, in some cases, the number of {\lq missed\rq} elements starts to rise before the number of {\lq false\rq} elements has dropped to zero, or, in other cases, the minimum number of one or both of these quantities is greater than zero. In such cases we have to automatically locate the best compromise for $\alpha$, such that it minimizes simultaneously both numbers. In order to define this optimum values of $\alpha$ we fit polynomials to the two sets of elements (see Fig.\ref{fit}) and we select as such the $\alpha$ value for which the fitted polynomials intercept. 

\subsubsection{Parameter $k$} 
\begin{figure}
\includegraphics[scale = 0.7]{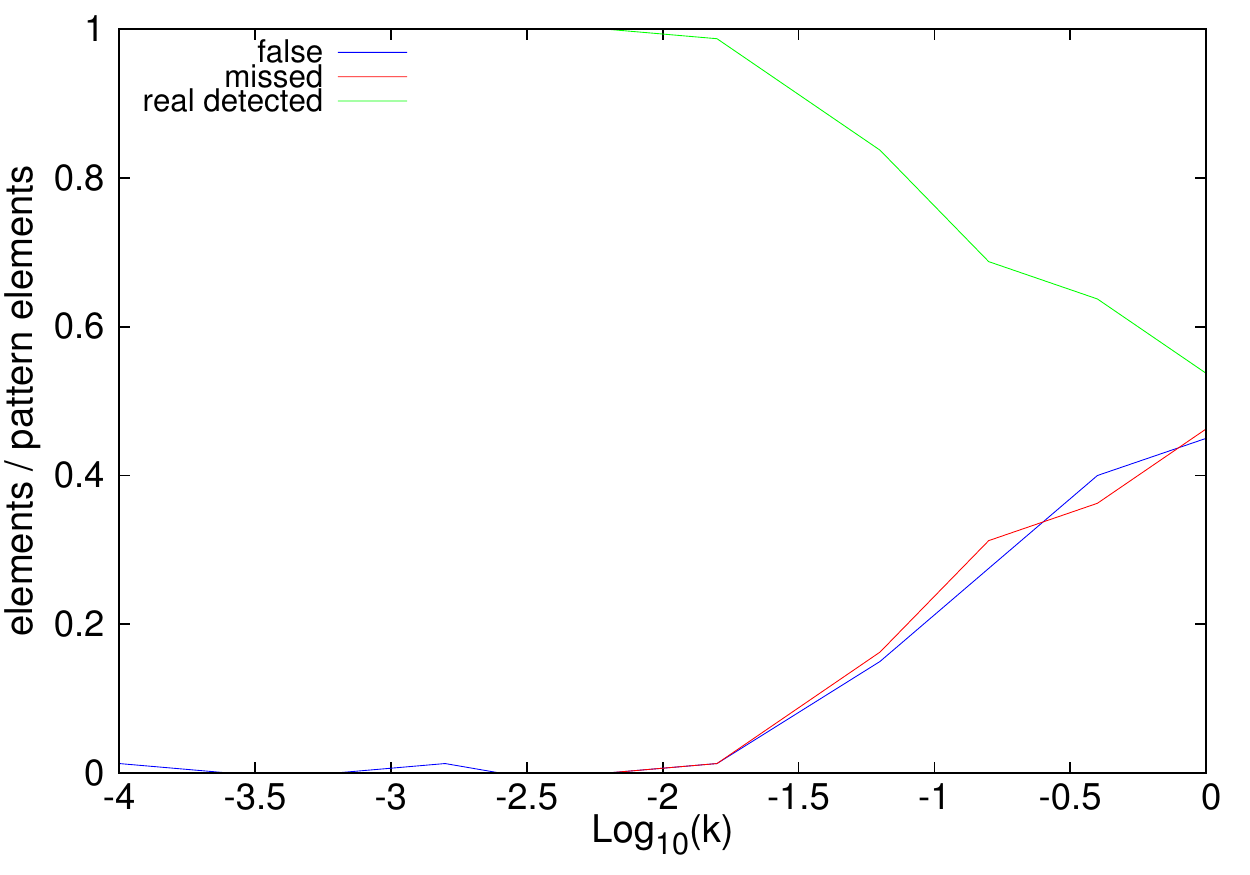}
\caption{Quality of the results of our algorithm as a function of $k$. The green line corresponds to the fraction of correctly detected pattern elements, the red line to the fraction of pattern elements which are missed and the blue line to the fraction of those which are falsely characterized as pattern elements by the algorithm. Since the number of {\lq real detected\rq} elements is equal to $1 - $ number of {\lq missed\rq} elements, we present from now on only the number of {\lq real detected\rq} elements. }\label{k_results}
\end{figure}

This parameter is introduced in our analysis via the definition of the {\em cocircularity coefficient} (see equation \ref{cocirc}). We wish to study how this parameter affects the success of the algorithm and also whether the two parameters, $\alpha$ and $k$, are interrelated. In Fig. \ref{k_results} we present the performance of the method (i.e. the fraction of {\lq missed\rq} elements in red, of false detections in blue and of real pattern-element detections in green) as a function of k. Note that for every value of $k$, the $\alpha$ parameter space is scanned in order to find the optimum $\alpha$-value, and the results shown are based on the individual optimum $\alpha$-values which are found for the specific $k$-value indicated.

\begin{figure}
\includegraphics[scale = 0.7]{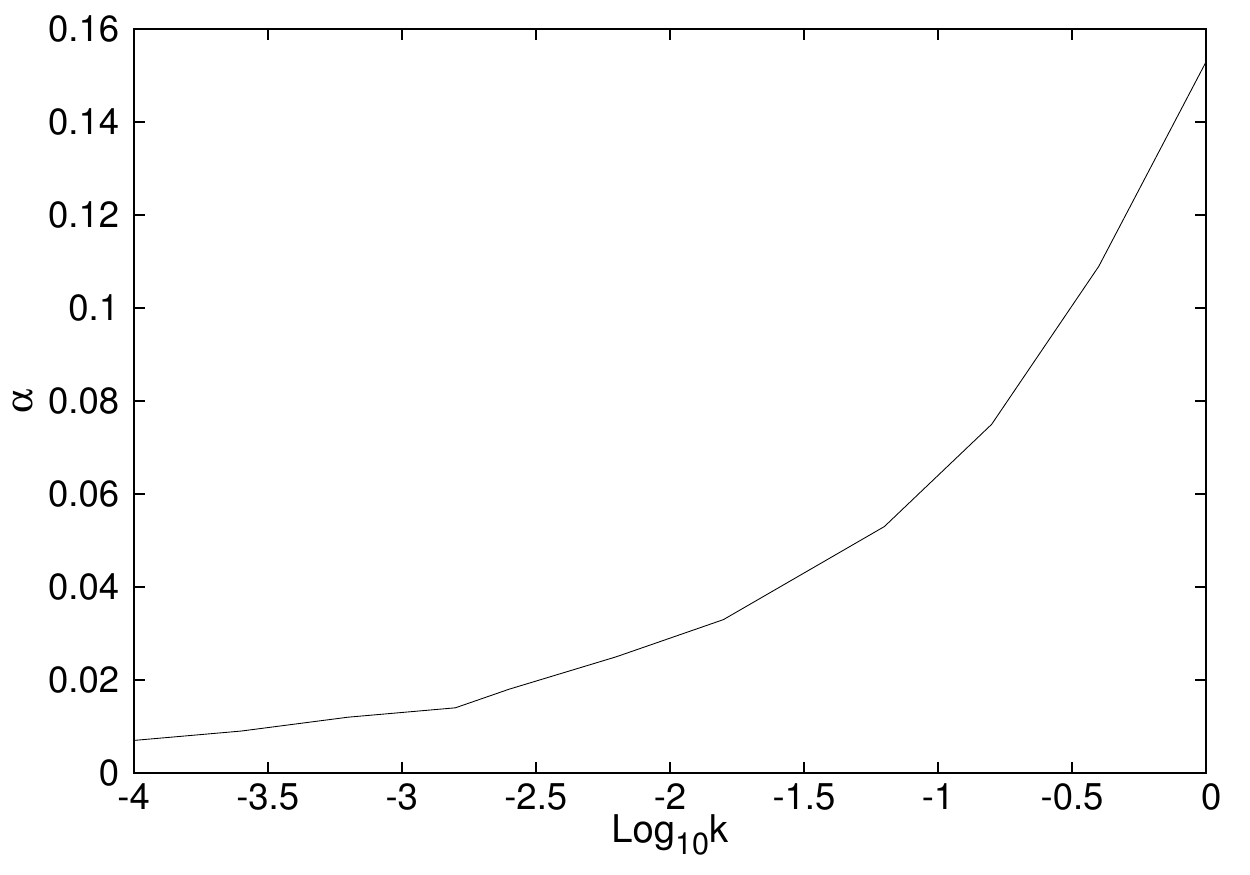}
\caption{The optimum value of $\alpha$ as a function of k for the image of Fig.\ref{pat440}.}\label{k_a}
\end{figure}

It is obvious that the qualitatively optimum results are obtained for small values of $\log(k)\leq -1.8$: both false and missed detections remain below $3$ per cent, i.e., $97-100$ per cent of the pattern elements are correctly detected. In Fig. \ref{k_a} we present the behavior of the optimum $\alpha$-value that is found as a function of $k$. Such a quasi-exponential shape of the $k(\alpha)$ curve is generic to all our test-bed images.

\subsubsection{Additional coefficients}
In the previous analysis, the interactions of the image elements were calculated by taking into account only their spatial separation and their cocircularity, i.e. whether their main axis is tangent to the same circle or curve. These are the simplest geometrical criteria for the detection of patterns, as those shown in Figure \ref{pat440}. More criteria can be inserted in order to improve the detections of patterns with specific properties, geometrical or other. Here we present two such examples. However, since the pattern recognition results on the image of Fig. \ref{pat440} are excellent, no improvement can be expected by using extra coefficients. Generally in Astronomy we require pattern recognition in images with significantly lower SN ratios. So, for a meaningful comparison of the results, when extra individual coefficients are taken into account, we have reduced the SN ratio of the image of Fig. \ref{pat440}, by reducing the density of the pattern in half, and we repeat the analysis on the new image constituted by $360$ noise elements and $40$ pattern elements (test-bed tag: $\rm P2$).

{\em Smoothness coefficient:} In order to favor the detection of smooth curves, which is the case, for example, of filamentary structures, we insert in the calculation of the interactions the {\em smoothness coefficient}, as it is defined in Section \ref{smooth_definition}, and we repeat the analysis on the new pattern.   

\begin{figure}
\centering
\includegraphics[scale=0.7]{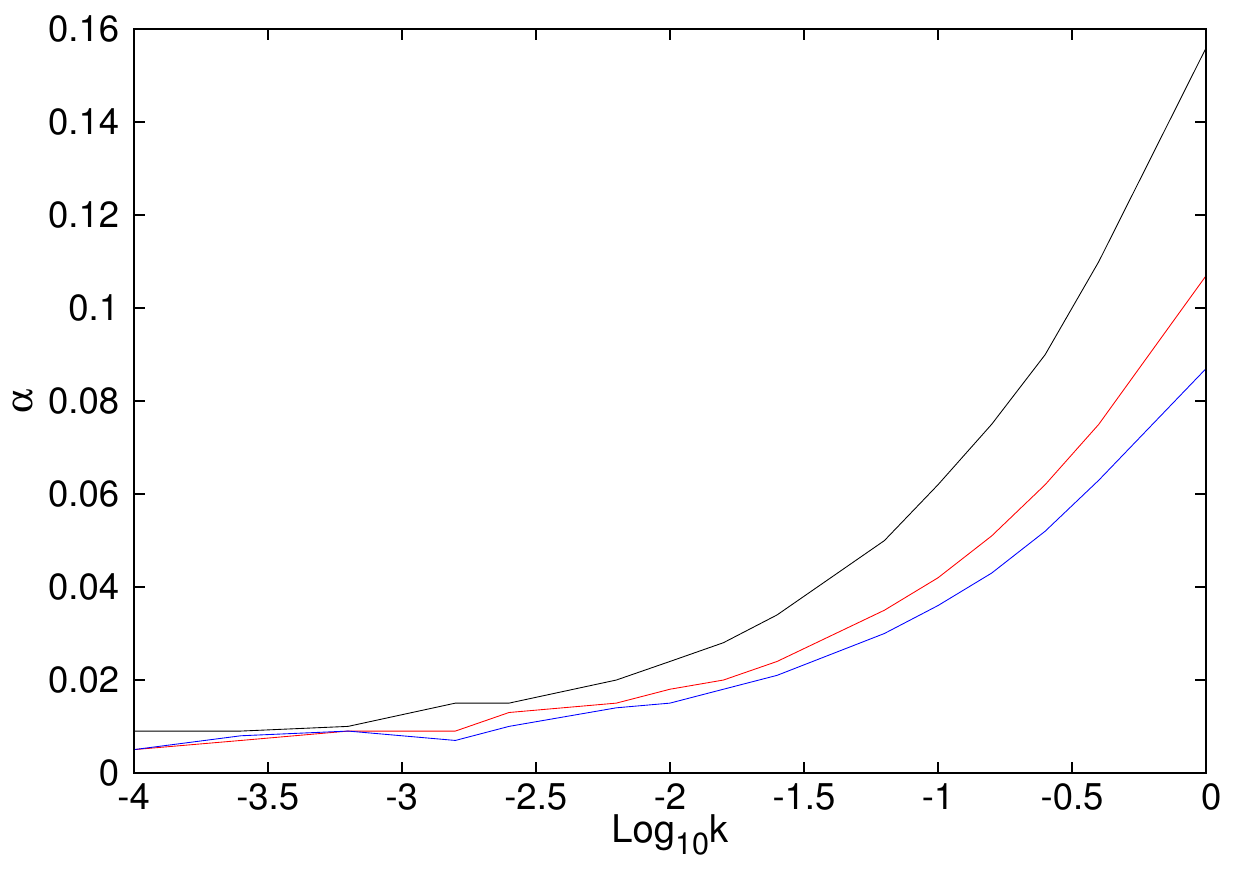}
\caption{The optimum values of $\alpha$ as a function of $k$ when {\em smoothness} (red curve) and also{\em mass} (blue curve) are taken into account for the calculation of the interaction coefficients compared with the corresponding curve of 
the simplest case using two individual coefficients, {\em proximity} and {\em cocircularity} (black curve).} \label{k_a_comp}
\end{figure}

First, we examine the effect of the extra coefficient on the $k - \alpha$ relation. In Fig. \ref{k_a_comp}, the relevant curves based only on {\em proximity} and {\em cocircularity} are shown in black and on {\em proximity}, {\em cocircularity} and {\em smoothness} in red. As it is evident, their quasi-exponential shape is similar in both cases, a fact found in every case that we have examined so far, but the values of $\alpha$ are systematically lower and their range for $k\in(0,1]$ narrower. This is the result of the nature of the $\alpha$ parameter itself, which is related to the SN ratio, or in other words, it is related to the number of elements that the algorithm is forced to identify as pattern elements. Thus, the more criteria we insert via the interaction coefficients, the better specified is the kind of pattern that the algorithm seeks to identify and, thus, the algorithm is less sensitive to the value of $k$ inserted in the definition of the {\em cocircularity} coefficient.

\begin{figure}
\includegraphics[scale=0.7]{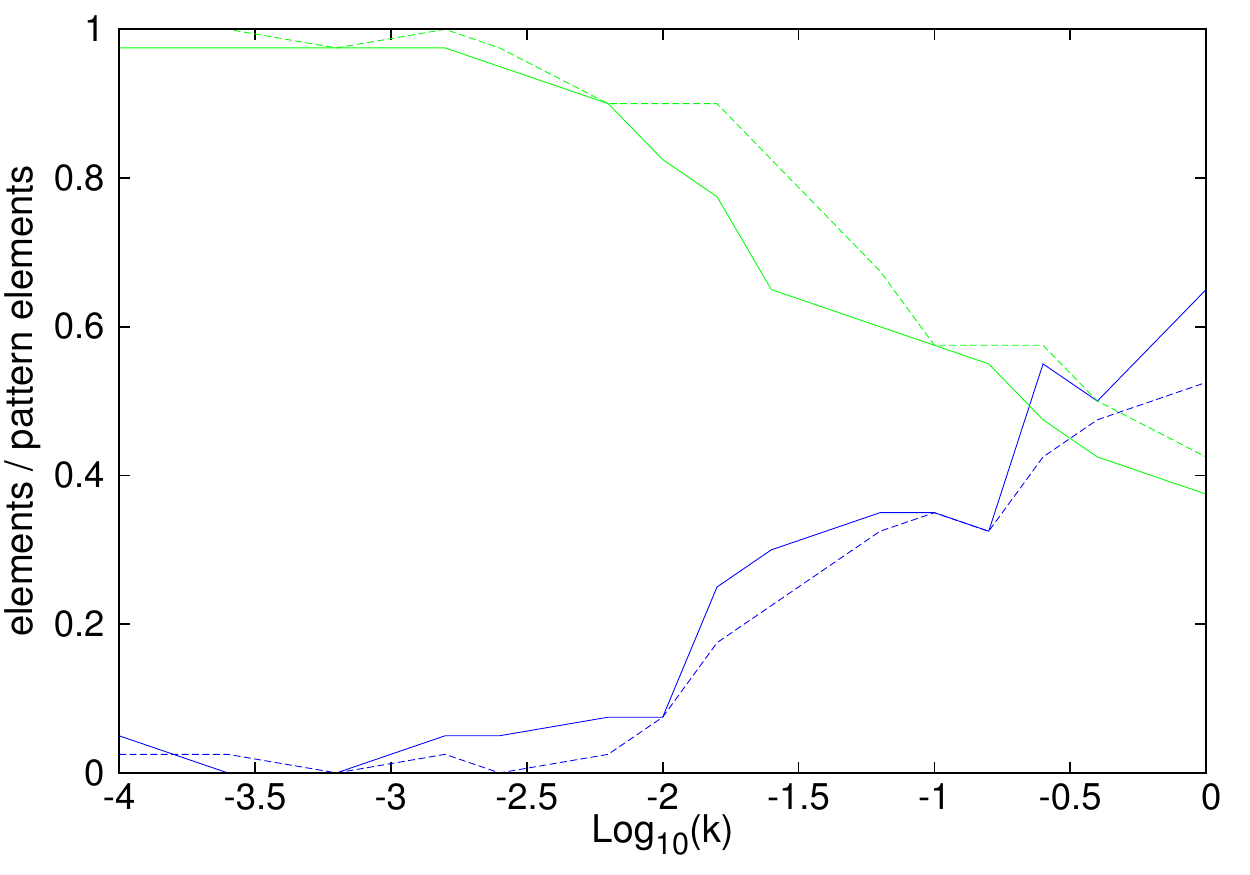}\hfill
\includegraphics[scale=0.7]{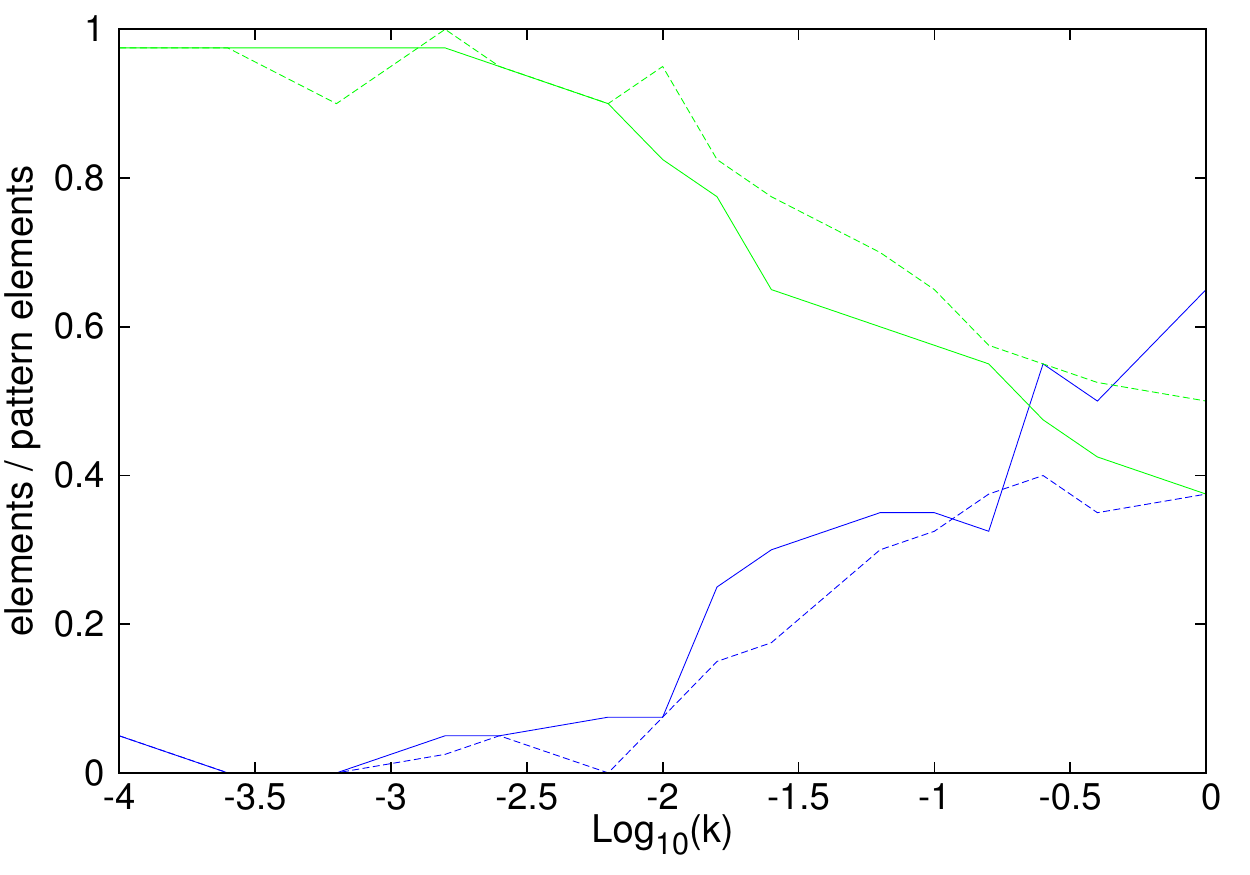}
\caption{Comparison of the detection quality as a function of the $k-$value when {\em proximity} and {\em cocircularity} are taken into account (solid lines) and when the {\em smoothness} coefficient is additionally taken into account (dashed line) (upper), and when the {\em mass coefficient} (dashed line) is further taken into account (lower). The green lines correspond to the fraction of image elements which are correctly tagged as {\lq pattern elements\rq} and the blue line to the number of noise-elements which are falsely characterized as {\lq pattern elements\rq}. All values are normalized to the total number of pattern elements.}\label{k_results_comp}
\end{figure}

Comparing the results shown in Fig. \ref{k_results_comp} ({\em upper panel}), we find that adding the criterion of {\em smoothness} (dashed lines) in the list of interaction coefficients qualitatively improves the detection. In most cases the number of the {\lq real detected\rq} elements (green) increases, while the number of {\lq false\rq} (blue) elements decreases significantly. Especially for intermediate $k$-values, for which the success of the algorithm is mediocre, the improvement increases by $12-13$ per cent. 

{\em Perturbing the orientation of the image elements:} The definition of the {\em smoothness coefficient} itself (see Section \ref{smooth_definition}) should enable one to detect patterns consisting of image elements, the orientation of which, is not exactly tangent to the curve formed by their positions, which should be the usual case for real patterns found in nature. Thus, we perturb the angles of the pattern of the image in Fig. \ref{pat440}, for three cases, by a random number between $\pm 5 \deg$, $\pm 10 \deg$, $\pm 20 \deg$ obtaining excellent detections for the first two cases (up to $99\%$ and $91\%$ respectively) and quite good results for the latter case (of the order of $75$ per cent). The range of $k$-values for which we obtain the best results have also a lower limit, since a very {\lq strict\rq} {\em cocircularity} criterion would not be satisfied by the perturbed orientations of the image elements and they would falsely be characterized as noise. The appropriate ranges are: $-2.8<\log{k}<-1.8$, $-2.2<\log{k}<-1.6$, $-2.2<\log{k}<-1.$, respectively.

{\em Mass coefficient:} The versatility of the method, presented in this paper, stems from the fact that one can insert new criteria, via individual interaction coefficients, in order to favor the patterns of interest, having specific geometrical or other physical characteristics. Thus, apart from the geometrical coefficients tested and proposed by \citet{HH1993}, we test one more coefficient, as an example of how one could adapt the method depending on the purposes of their specific application. For the current example we apply our algorithm on N-body simulation data, adding an extra {\em mass coefficient}, since dynamically bound structures, which correspond to deep potential wells, host massive DM haloes and thus, the relatively massive image elements could be used to trace structure. Another alternative observable could be the luminosity or colour of galaxies (see Section \ref{mass_definition}).  

For the application of the algorithm using the {\em mass coefficient}, we assigned a mass to each point of the test-bed image, so that it follows the AF which we extracted from the $\Lambda$CDM DEUS simulated light-cone halo catalogue data in \cite{Chira2018}. We allowed noise points to take random values from the whole mass range of the AF, with $M \geq ~10^{13} M_{\odot}$, while to each pattern point, were assigned randomly a mass from the AF with $ M \geq 10^{14} M_{\odot}$. 

In Fig. \ref{k_a_comp}, the blue curve corresponds to the use of all four interaction coefficients ({\em proximity, cocircularity, smoothness, mass}) and we see that the extra {\em mass coefficient} has a measurable effect on the range of the optimum values of $\alpha$, as a function of k, 
indicating that using more interaction coefficients renders the optimum $\alpha$ values less dependent on the value of $k$. 

In Fig. \ref{k_results_comp} we compare the quality of the pattern detection, as a function of the $k$-value, for three cases where we use an increasing number of interaction coefficients, ie.,
(1) the {\em proximity} and {\em cocircularity} coefficients (solid lines), (2) adding the {\em smoothness} coefficient (the dashed lines in upper panel) and (3) adding both the {\em smoothness} and the {\em mass} coefficients (the dashed lines of lower panel). With increasing number of interaction criteria, we obtain an increasing (decreasing) number of {\lq real detected\rq} ({\lq false\rq}) pattern elements for $\log{k} \gtrsim -2.5 $, while no significant difference is found for $\log{k}\lesssim-2.5$. The cause of the latter behaviour is the exact {\em cocircularity} of our mock patterns in our $\rm P2$ test-bed image, i.e. for small values of the $k$ parameter (strict {\em cocircularity} coefficient), the contribution of the extra coefficients to the detection performance is very small or even negligible. However, the improvement for intermediate values of the $k$ parameter shows that characterizing patterns, via multiple, meaningful coefficients, is very important for the successful application of the algorithm on images of unknown properties, for which the exact optimum values of the parameters are unknown. 

We note here that excluding the {\em smoothness} coefficient, the effect of the {\em mass} coefficient becomes more prominent, improving the pattern detection performance, in a similar fashion as that of the {\em smoothness} coefficient.

\subsection{Performance as a function of the SN Ratio} 
\begin{figure}
\centering
\includegraphics[scale = 0.7]{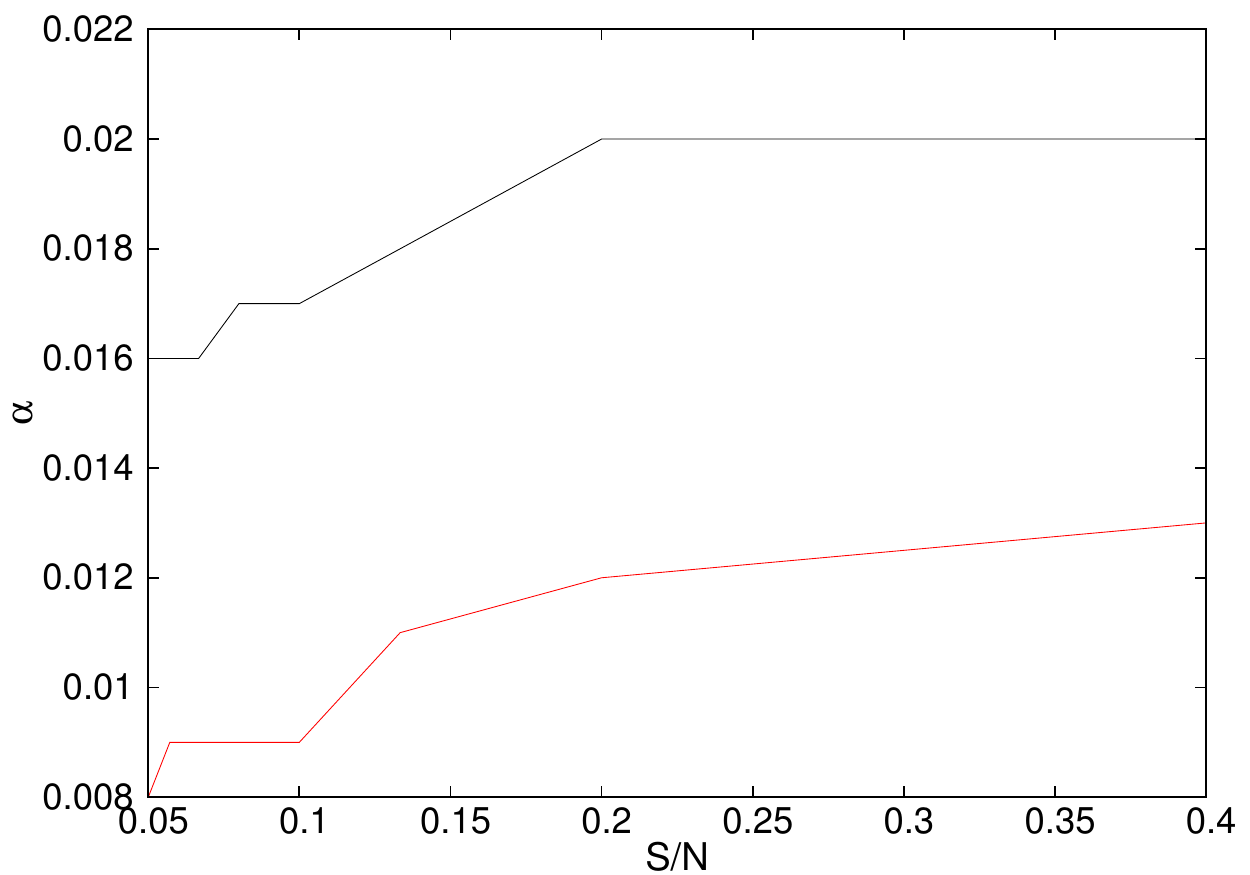}
\caption{The optimum value of $\alpha$ as a function of the SN ratio. For $\log{k}=-2.8$ (red curve) and $\log{k}=-2.$ (black curve).}\label{sn_a}
\end{figure}

The $\alpha$ parameter is by definition strongly correlated to the SN ratio, but the exact relation of the optimum $\alpha-$value to the SN ratio is rather unknown. Also, it would be interesting to clarify if and to what extend the value of $k$ affects the optimum $\alpha-$SN relation. For this purpose, we construct a grid of images with different SN ratio containing the same, low-density, pattern of 40 points, with the SN ratio ranging between $0.05$ and $0.4$. We present in Fig. \ref{sn_a} the relevant results based on the {\em proximity}, {\em cocircularity} and {\em smoothness} coefficients for two different values of $k$: $\log{k}=-2$ and $\log{k}=-2.8$ (red), which in Fig. \ref{k_results_comp} leads to optimum results. 

\begin{figure}
\centering
\includegraphics[scale = 0.7]{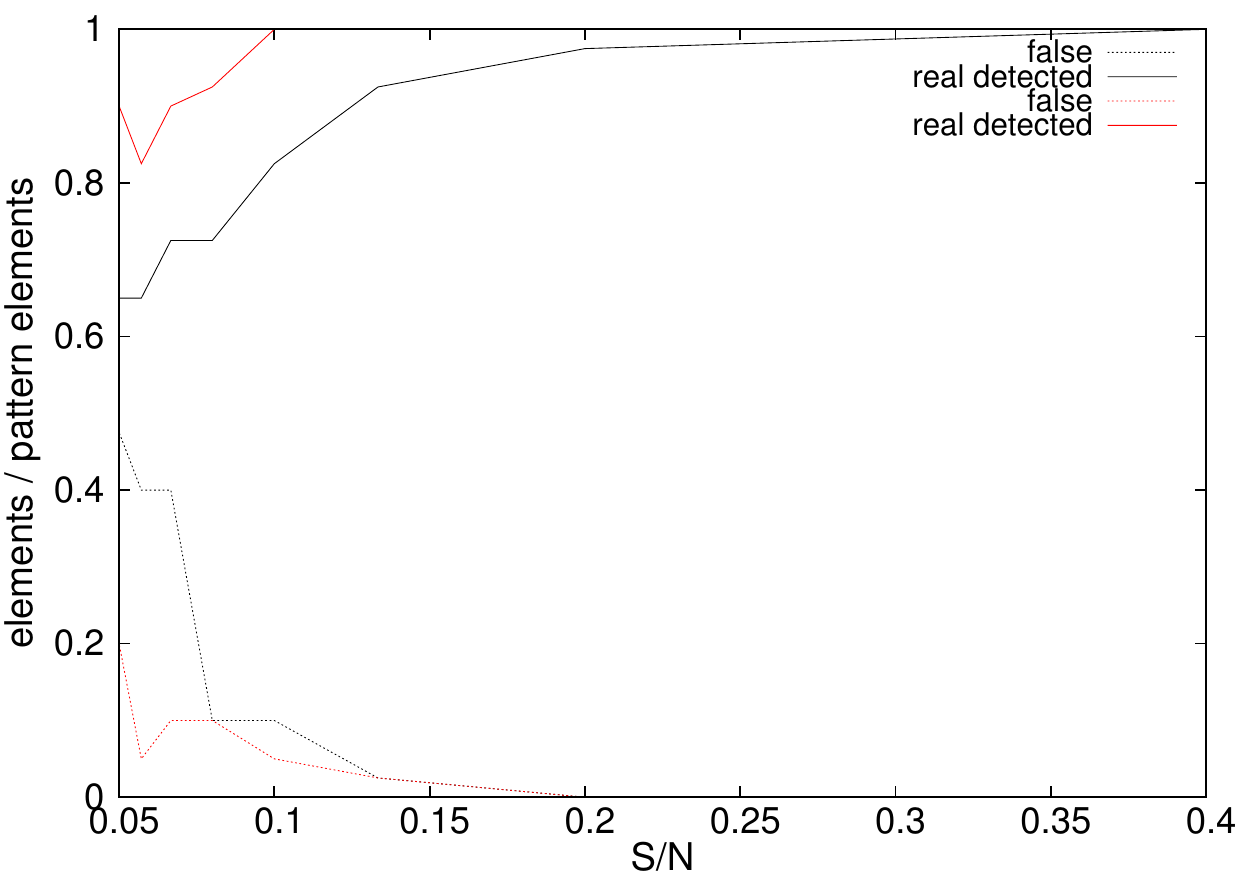}
\caption{The results obtained for different SN ratio for $\log{k}=-2.8$ (red curves) and $\log{k}=-2$ (black curves). The different quantities of interest are shown in different line-styles as indicated.}\label{sn_results}
\end{figure}

Fig. \ref{sn_a} shows that the optimum value of $\alpha$ has a  weakly increasing trend with the SN ratio, although the variation range is quite small. Also, in Figure \ref{sn_results} it is obvious that the success of the algorithm drops with decreasing SN ratios, as expected. However, the performance is significantly better for the optimum value of $\log{k} =-2.8$. We note that for the majority of the SN values the fraction of {\lq false\rq} and {\lq missed\rq} elements decreases by more than $\sim 50$ per cent for the lowest $SN$-values when $\log{k}$ is decreased from $-2$ to $-2.8$. However, we also note that, down to SN$=0.2$, $\log{k}=-2$ is also a very good choice. 

\begin{figure*}[p]
\captionsetup[subfigure]{labelformat=empty}
\centering
    \begin{subfigure}[b]{0.475\textwidth}
    \centering
    \includegraphics[width=0.93\textwidth]{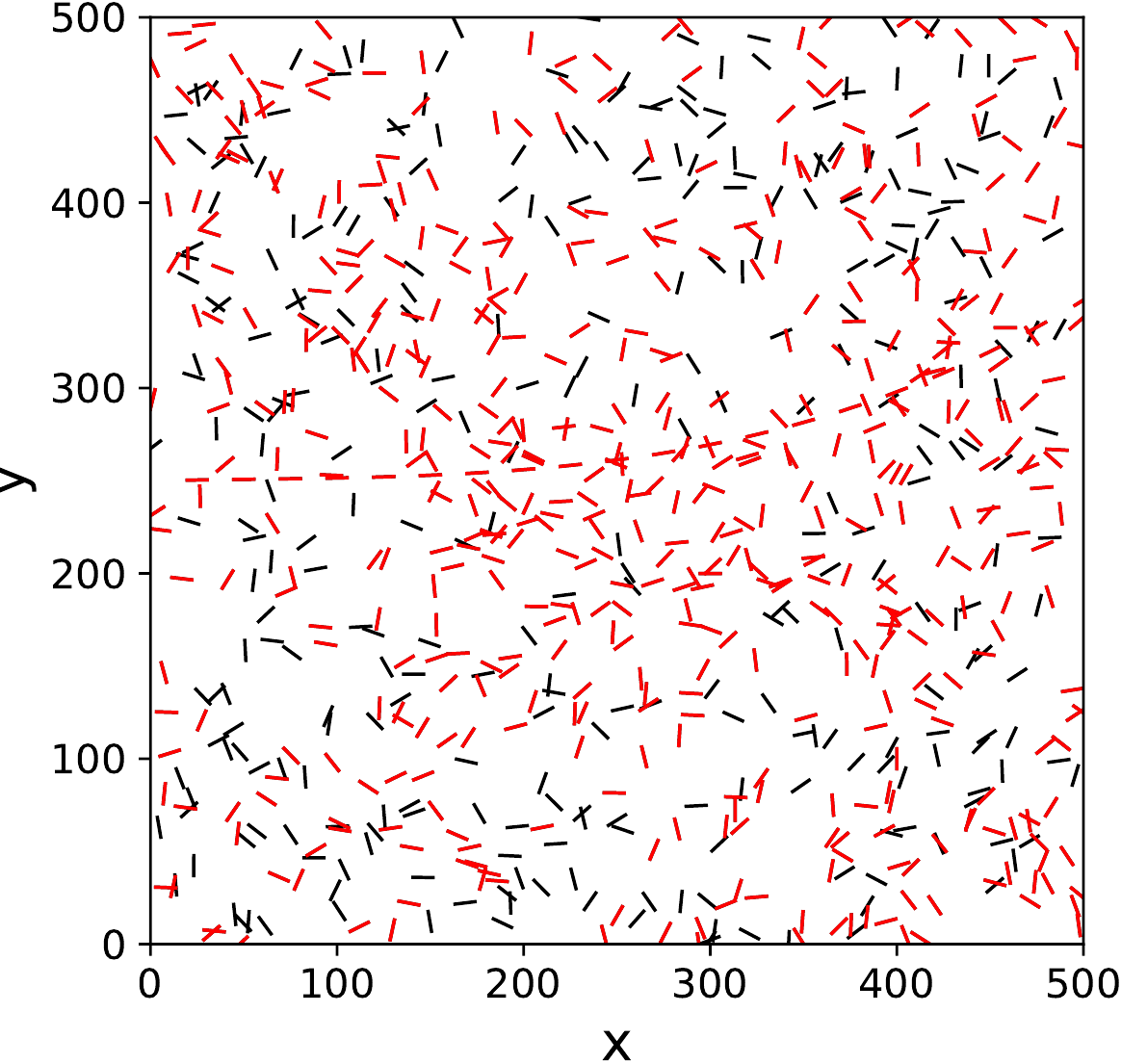}
    \end{subfigure}
    \hfill
    \begin{subfigure}[b]{0.475\textwidth}
    \centering
    \includegraphics[width=0.93\textwidth]{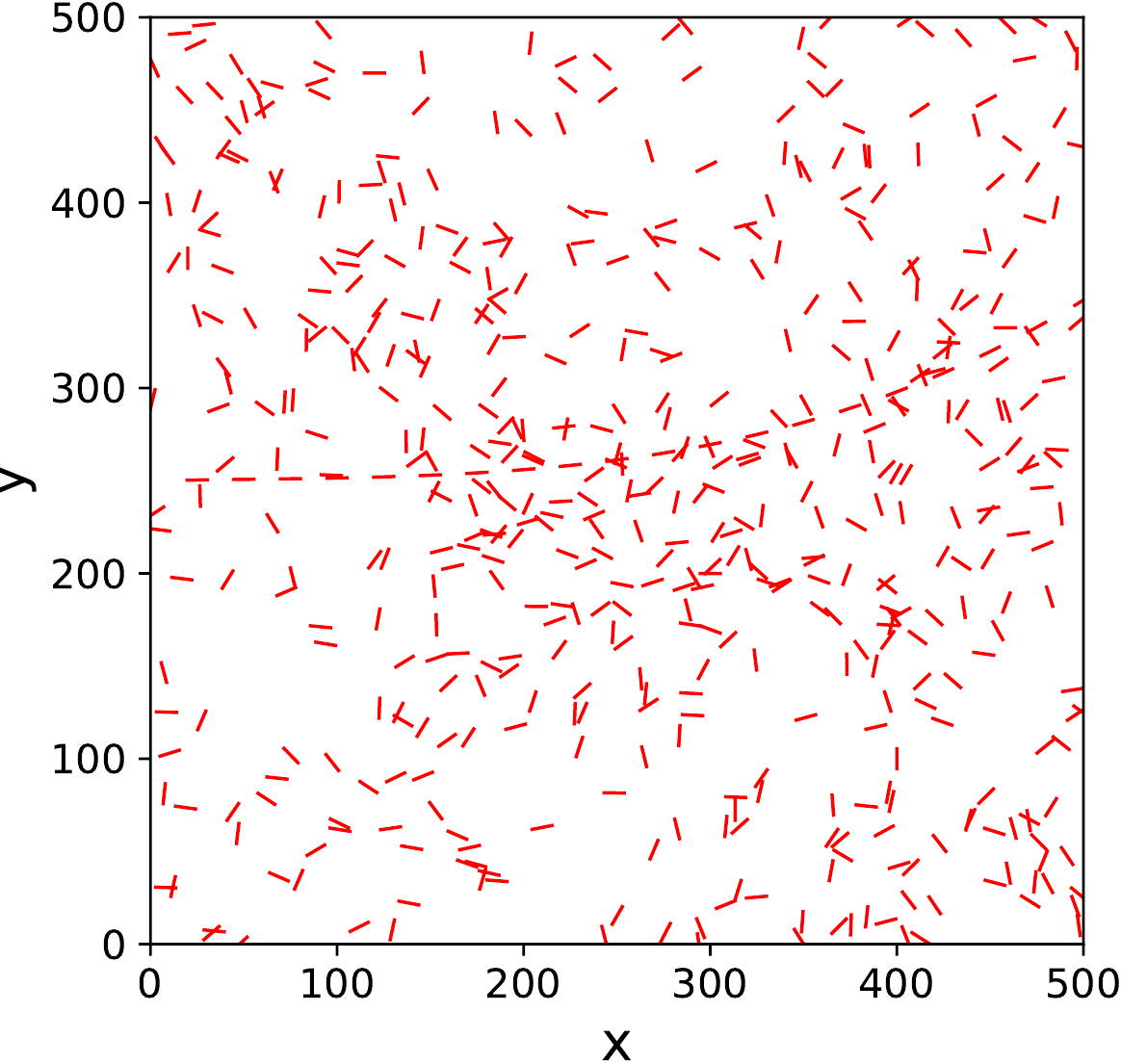}
    \end{subfigure}
    \hfill
    \begin{subfigure}[b]{0.475\textwidth}
    \centering
    \includegraphics[width=0.93\textwidth]{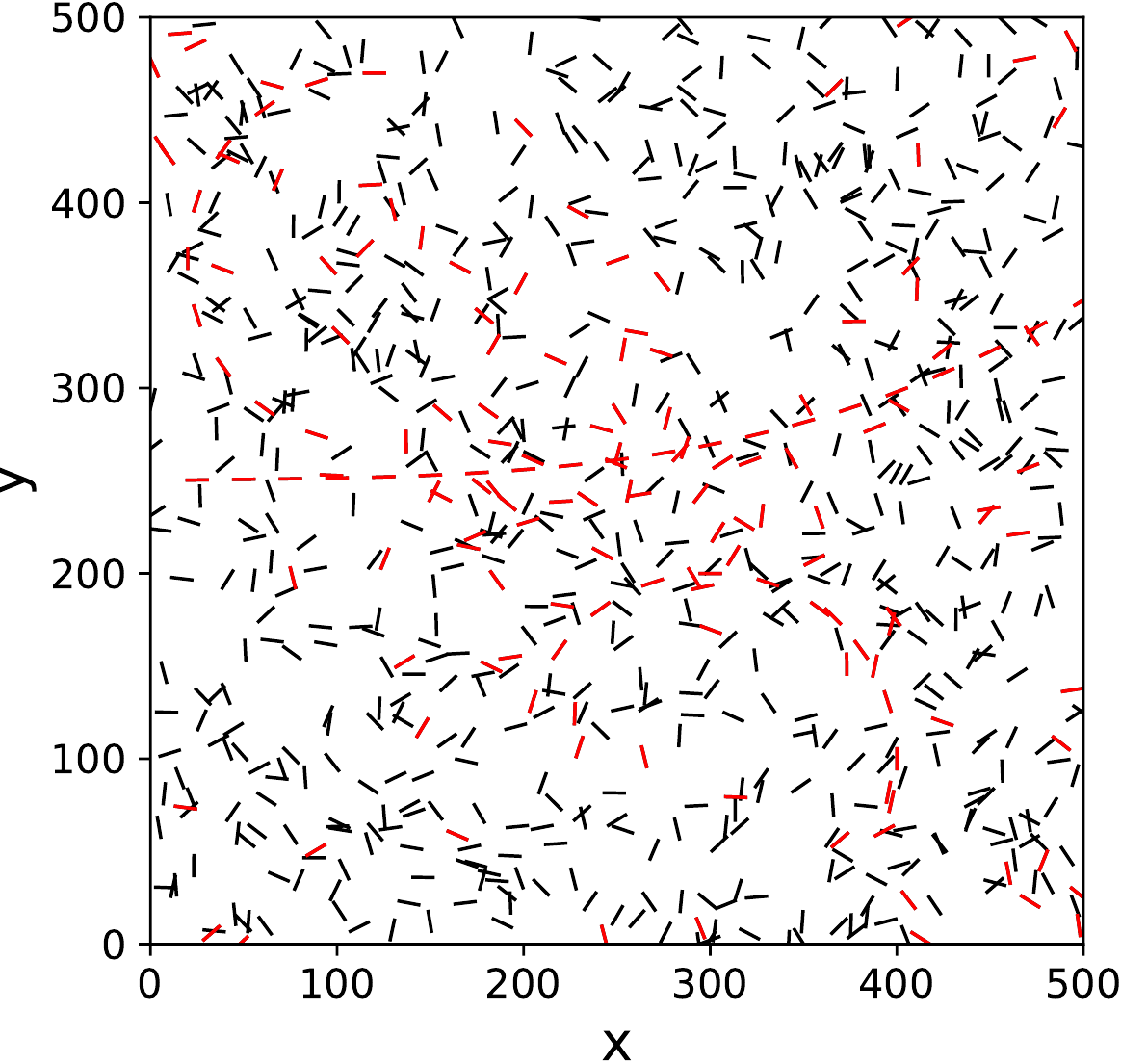}
    \end{subfigure}
    \hfill
    \begin{subfigure}[b]{0.475\textwidth}
    \centering
    \includegraphics[width=0.93\textwidth]{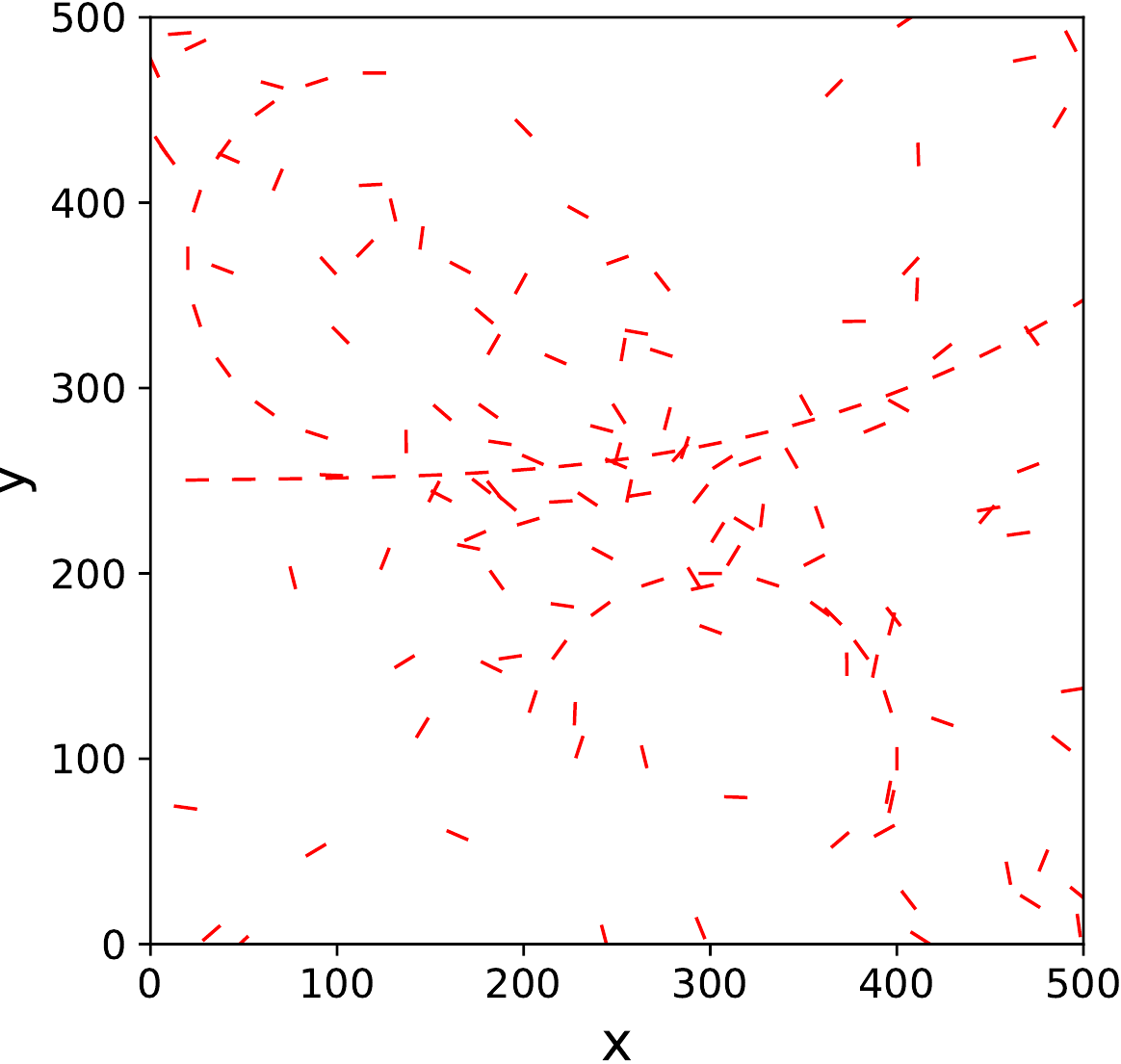}
    \end{subfigure}
    \hfill
    \begin{subfigure}[b]{0.475\textwidth}
    \centering
    \includegraphics[width=0.93\textwidth]{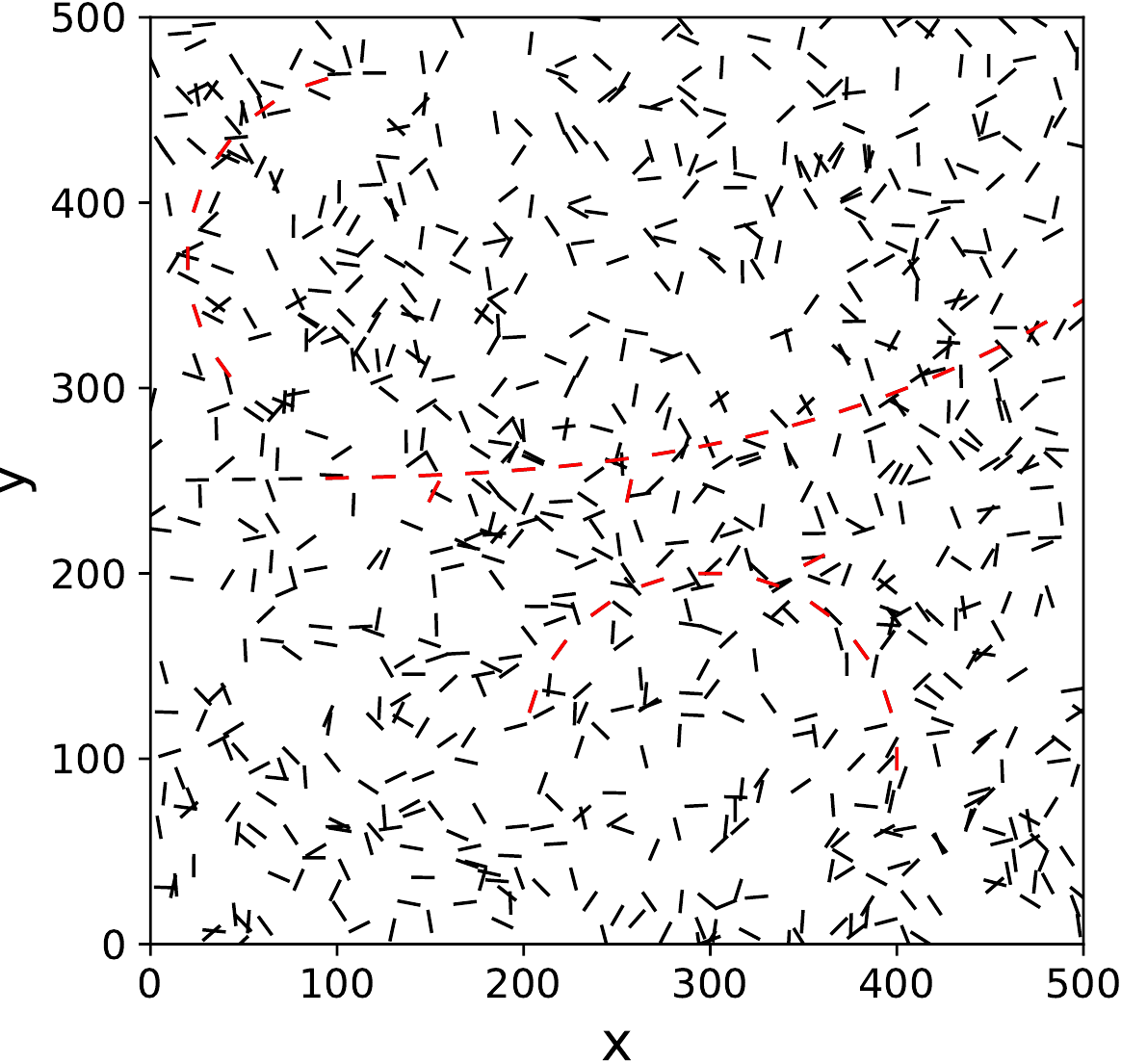}
    \end{subfigure}
    \hfill
    \begin{subfigure}[b]{0.475\textwidth}
    \centering
    \includegraphics[width=0.93\textwidth]{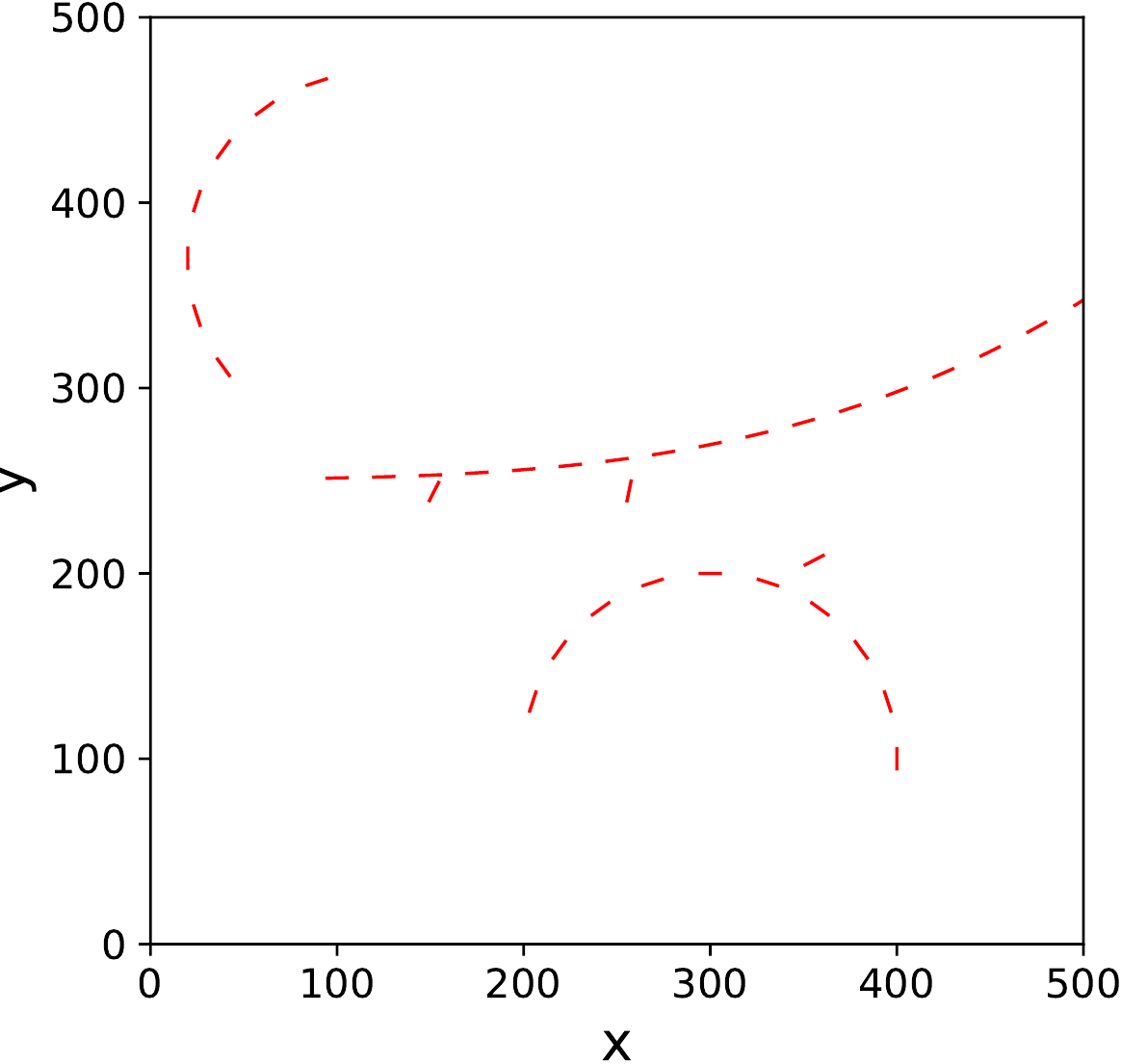}
    \end{subfigure}
\RawCaption{\caption{ The results of pattern-recognition for SN$= 0.05$, for sequential values of $\alpha$ up the optimal. The image consists of 40 pattern elements and 800 noise elements. The value of $\alpha$ increases from left top to bottom right. The detected pattern elements are shown in red on the image (left column) and separately (right column). The optimum detection (bottom) corresponds to 3 {\lq false\rq} and 6 {\lq missed\rq} elements.}\label{fig:sequence}}
\end{figure*}

In order to visualize the pattern detection procedure as a function of $\alpha$, we present in Fig. \ref{fig:sequence} the results of our algorithm for the case of the lowest SN ratio image, $SN=0.05$, consisting of 40 pattern elements and 800 noise elements. The sequence starts at the upper row of the plot and reaches the optimum value of $\alpha$ at the lower row. In order to enhance visually the success of the algorithm, we present at the right-hand panels of the figure only the elements detected as pattern by our algorithm, while at the left-hand panels we present all the image elements (detected as noise in black, detected as pattern in red). As it can be seen the detection results for the optimum case are quite impressive.

\subsection{An application to gravitational lensing detection.}

\begin{figure*} \centering
    \begin{subfigure}[b]{0.45\textwidth}
        \includegraphics[width=\textwidth]{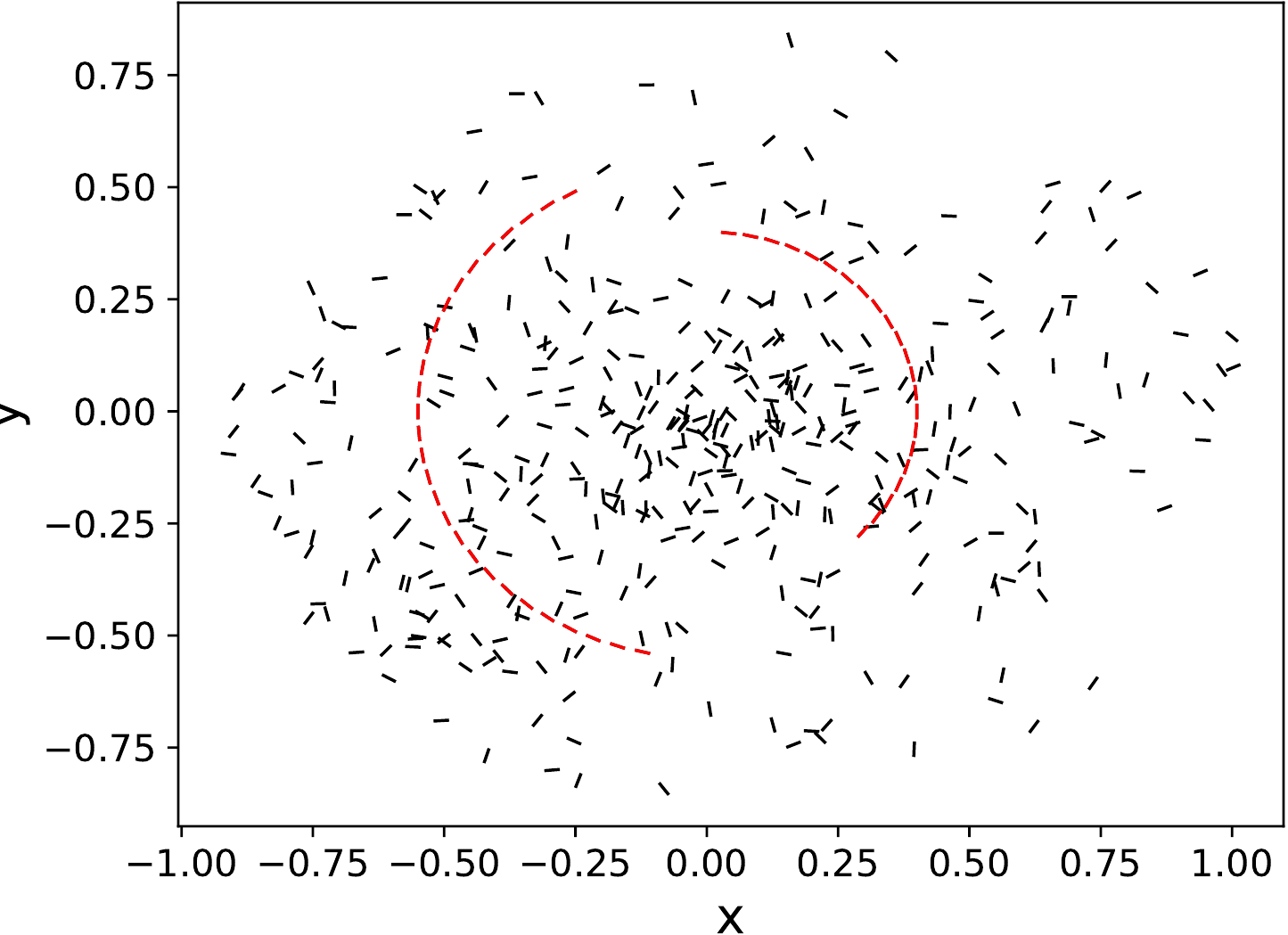}
        \label{cluster:a}
    \end{subfigure} %
    \begin{subfigure}[b]{0.48\textwidth}
        \includegraphics[width=\textwidth]{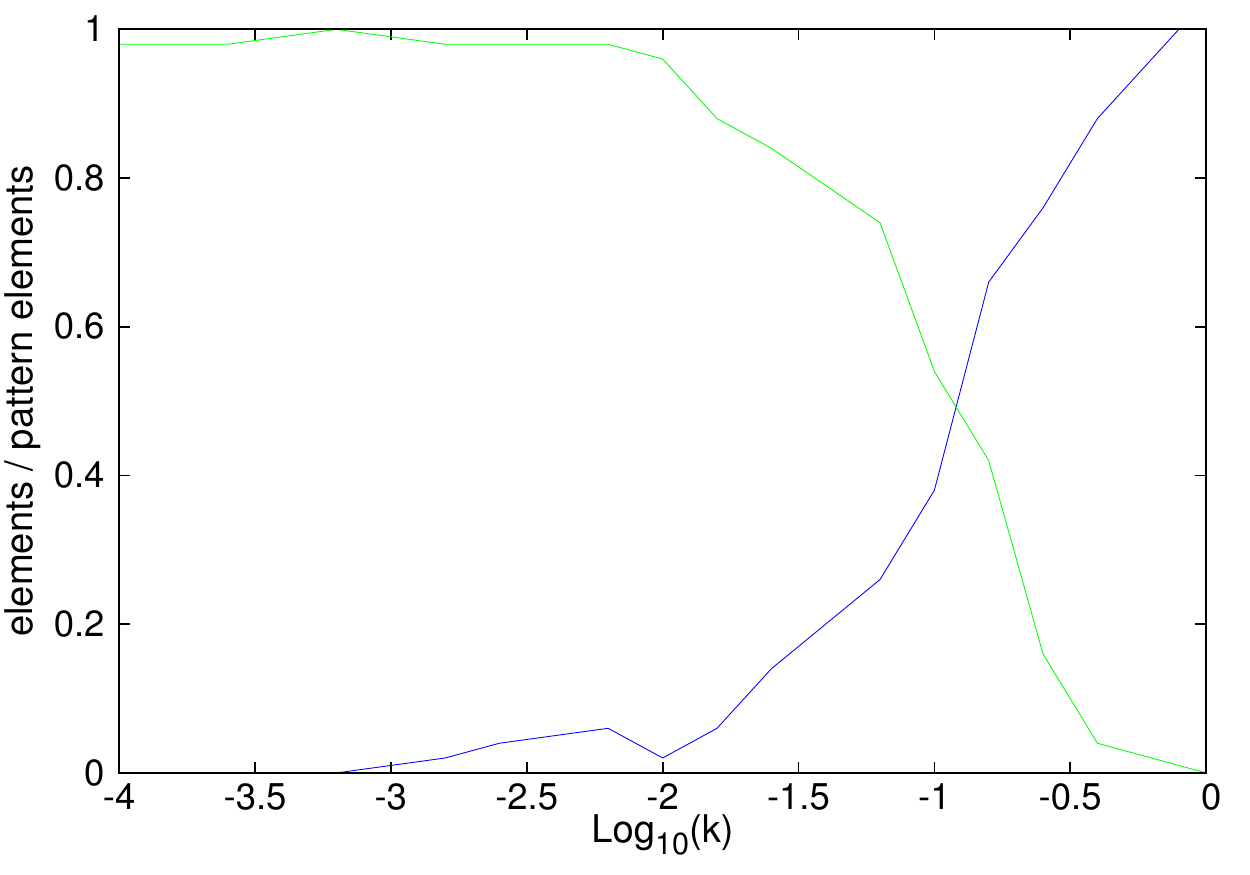}
        \label{cluster:c}    
    \end{subfigure}
    \ \\
    
    \begin{subfigure}[b]{0.45\textwidth}
        \includegraphics[width=\textwidth]{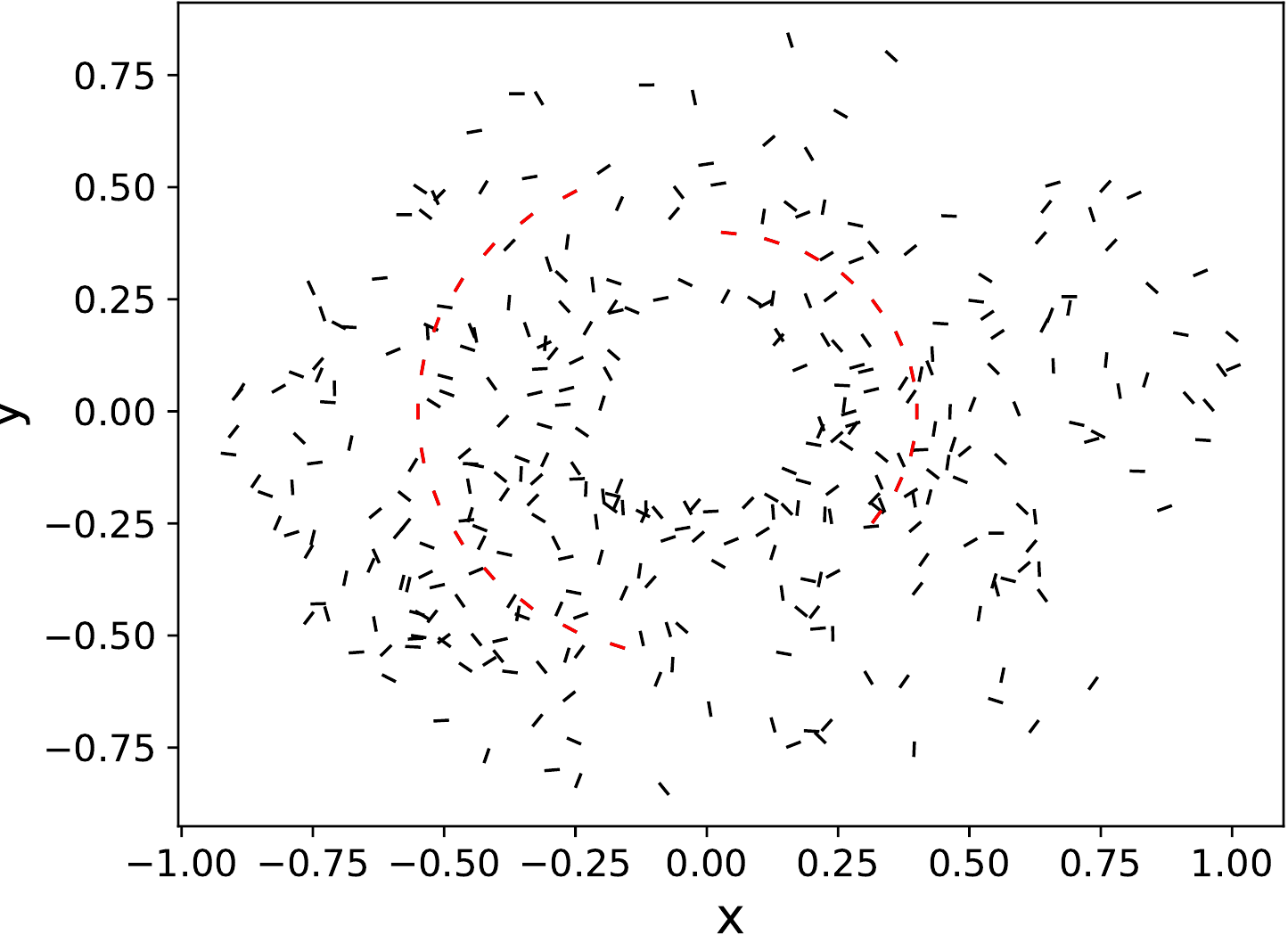}
        \label{cluster:d} 
    \end{subfigure}
    \begin{subfigure}[b]{0.48\textwidth}
        \includegraphics[width=\textwidth]{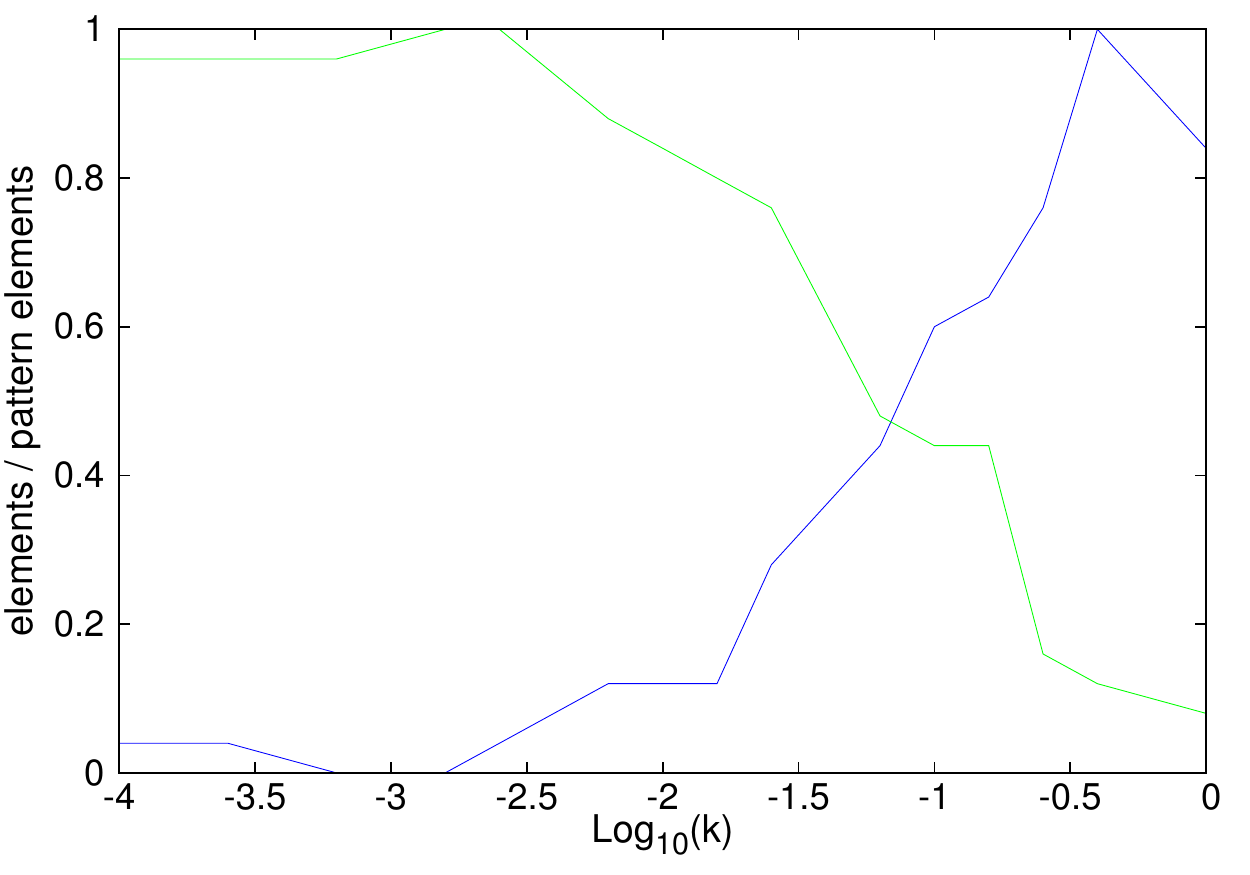}
        \label{cluster:b}    
    \end{subfigure} 
    \caption{Images simulating the effects strong gravitational lensing. Cluster {\em Abell 1656} consisting of 419 galaxies. Top left:  two arcs at $R=0.4 R_{\rm cluster}$ and $R=0.55 R_{\rm cluster}$ constituted of 50 elements (test-bed tag: $\rm CL1$). Bottom left: Pattern number density decreased to half (25 pattern-elements)-subtracted center of cluster (test-bed tag: $\rm CL2$). On the right of each image is presented the respective detection-quality as a function of the $k$-value. The green lines correspond to the image elements which are correctly tagged as {\lq pattern elements\rq} and the blue line to the number of noise elements which are falsely characterized as {\lq pattern elements\rq}. All values are normalized to the total number of pattern elements.}\label{cluster2}
\end{figure*}

So far, we have worked on mock images of different patterns representative of the large-scale structure of the Universe, like curves and lines, to find that our algorithm identifies successfully a large fraction of the patterns for specific values of the parameters $\alpha$ and $k$. In this paragraph we approach a different class of images in order to further investigate how the specific methodology can be used for other astronomical applications. Specifically, the lensing configuration resembling an Einstein's ring   \citep{Refsdal1994,Kochanek2001,Bartelmann2010} has geometrical characteristics that should render it detectable by our methodology. 
Thus, we investigate if it would be possible to use our methodology to automatically detect strong lensing phenomena in cluster images.

To this end, we used the galaxy distribution of the cluster {\em Abell 1656} consisting of 419 SDSS galaxies with r-magnitude $<17.77$ and we inserted various types of mock patterns resembling an arc distribution (as it would result from a spherical gravitational potential). As a characteristic example of our relevant applications, we present in Fig.\ref{cluster2} the test-bed images analyzed below. In detail we have applied the algorithm taking into account the criteria of {\em proximity, cocircularity} and {\em smoothness}. We have analysed a variety of configurations, with equivalent results, but chose to present here these representative applications, based on a pattern which consists of two large arcs at different radii with a total of 50 {\lq lensed\rq} elements.

The elements of the two circular arcs are distributed at a distance of $R=0.55 R_{\rm cluster}$ (left arc) and the  $R=0.4 R_{\rm cluster}$ (right arc). Note that the image elements belonging to the two different arcs are not exactly cocircular, and this short of {\lq discontinuity\rq} in the pattern complicates its detection. However, an excellent detection is achieved for $\log{k}\leq -1.8$ (see Fig.\ref{cluster2}, top-right), which corresponds to a range of optimum $\alpha-$values, $\alpha \in [0.012,0.027]$.

Going a step further in testing more realistic images, we reapply our algorithm after decreasing the density of the pattern by a half (see Fig. \ref{cluster2}, bottom). We also exclude the dense cluster core since we expect that the {\em proximity} coefficient could dominate the performance of the pattern recognition algorithm and the recovered pattern could be contaminated by elements of the cluster core. Therefore,  we subtract the dense core of the cluster ($R<0.2 R_{\rm cluster}$) and rerun our detection algorithm (see Fig.\ref{cluster2},bottom). The results show an excellent pattern detection for a much wider range of the $k$ parameter, $\log{k}<-1.2$ (see Fig.\ref{cluster2}, bottom-right panel).As shown in Fig.\ref{cluster2} (bottom right), we again excellently recover the pattern for the same range of $k$ ($\log{k}\leq -1.8.$) corresponding to a slightly reduced range of optimum $\alpha$ values, ie., $\alpha \in [0.011,0.024]$. Note that in spite of subtracting the central galaxies, the SN ratio in Fig.\ref{cluster2} (bottom panel) is still lower than that of Fig. \ref{cluster2} (top panel).

\begin{figure*}
\centering
\includegraphics[scale=0.9]{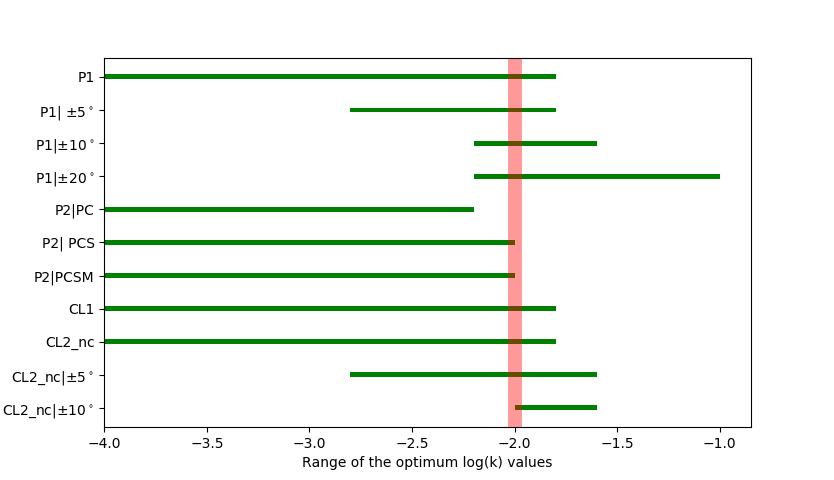}
\caption{The interval of optimum $k$ values for the different test-bed images used. From top to bottom: {\bf P1:} is based on the test-bed of Fig.\ref{pat440} (80 pattern elements and 360 noise elements), using the {\em proximity} and {\em cocircularity} interaction criteria. {\bf P1$\vert\pm 5$,$\pm 10$,$\pm 20$:} the orientations of the image elements of image P1 perturbed randomly within the denoted range. {\bf P2 $\vert$ PC:} The previous case image with the pattern number density reduced to half (40 pattern elements and 360 noise elements). {\bf P2$\vert$ PCS:} interaction criteria: {\em proximity},{\em cocircularity} \& {\em smoothness}. {\bf P2$\vert$ PCSM:} interaction criteria: {\em proximity},{\em cocircularity}, {\em smoothness} \& {\em mass}. 
{\bf CL1:} Test-bed of Fig. \ref{cluster2}(top). 
{\bf CL2$_{\rm nc}$:} Test-bed of Fig.\ref{cluster2}(bottom). 
{\bf CL2$_{\rm nc} \vert$ $\pm 5\deg$, $\pm 10\deg$:} in the image of Fig.\ref{cluster2}(bottom) the orientations of the image elements are perturbed randomly in the denoted range. 
The value $\log{k}=-2.$, pointed out with the vertical, red, thick line, represents the best compromise that suits the majority of test-beds.}\label{k_summary}
\end{figure*}

As a last, even more realistic application, we have perturbed randomly the orientations of the image elements for one of the previous cases, Fig. \ref{cluster2} (bottom panel), by $\pm 5 \deg$ and $\pm 10 \deg$, to obtain pattern detection up to $90$ per cent and $65$ per cent for optimum values of the $k$ parameter spanning a range of $(-2.8,-1.6)$ and $(-2.,-1.6)$, respectively. The respective optimum $\alpha$ ranges are $(0.010,0.029)$ and $(0.017,0.025)$. 
The fraction of {lq false\rq} elements is $\sim10$ per cent and $\sim35$ per cent for the previous cases, respectively, which due to the definition of our optimum detection is of the order of the {\lq missed\rq} elements.

This last test confirms that the current algorithm could be used for astronomical applications since it is able to detect patterns with statistical and not precise alignments. Thus, we move on applying our algorithm on a real optical cluster-image in a following section of this paper, making use of the optimum range of the parameters, as it is constrained in the analysis discussed so far. 

\subsection{Summary on the basic method}\label{partial_sum}
The essence of our analysis so far is mainly encapsulated in the intrinsic study of the free parameters that enter in the definition of main interaction coefficients, ie.,  $k$ and $\alpha$ and their effective range, in order to understand the role of these parameters and their effect on the success of the pattern identification.

Our most clear conclusions are those related to $k$, the parameter introduced in the definition of the {\em cocircularity coefficient}. Its value determines how strictly the pattern characterization depends on the criterion of {\em cocircularity}. For the case where the orientation of the pattern elements define exact circular patterns, the algorithm provides excellent recovery of the pattern for a wide range of $\log{k}\lesssim -2$. However, in more realistic cases, where the orientations of the pattern elements are  perturbed relatively to the tangent of the curve defined by their positions, the optimum interval for the $k$ parameter has a lower limit, the value of which increases with increasing perturbation amplitude. 

The former conclusions are summarized graphically in Fig. \ref{k_summary}, where we show in green bars the range of $k$-values that lead to optimum pattern detections for each of the test-bed images discussed in the current article (each test-bed image is denoted in the caption and given a specific tag-name). In the same plot, we see that the optimum values are found to be always $\log{k}\leq -1$ and that the value $\log{k}=-2$, pointed out with the vertical, red, thick line, is a very good compromise for the large majority of our images. 

Only for one of the cases presented here (P2$\vert$PC), the optimum $k-$value does not fall in the common range. This low-density pattern (40 pattern- out of 400 total image-elements), detected using only the {\em proximity} and {\em cocircularity} coefficients, demands a more strict {\em cocircularity} coefficient in order to be detected.

\begin{figure}
\centering
\includegraphics[scale=0.7]{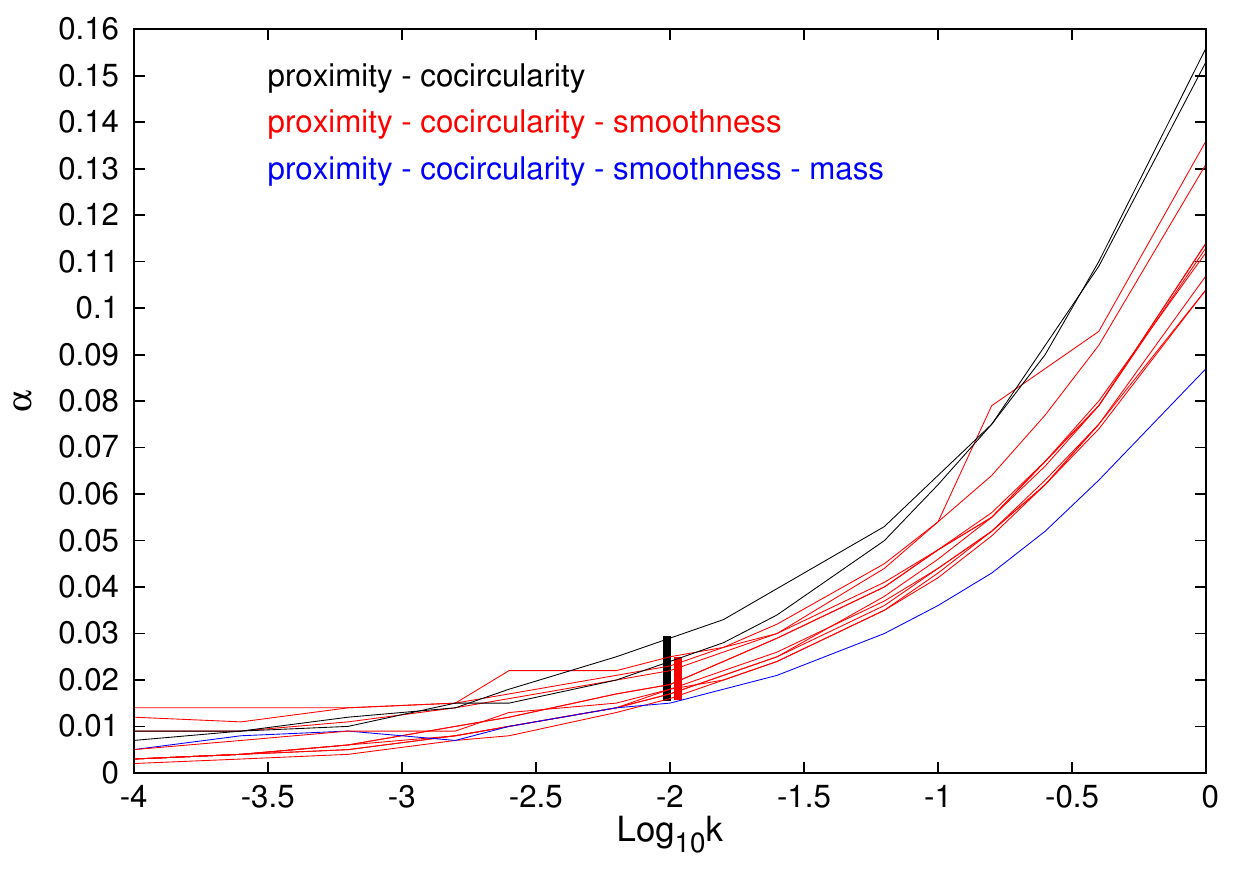}
\caption{The $\log{k}-\alpha$ curves for all the tests discussed in the current work. The results obtained only with the criteria of {\em proximity} and {\em cocircularity} are shown in thin black curves, those obtained with the criteria of {\em proximity}, {\em cocircularity} and {\em smoothness} are shown in thin red curves and the case for which {\em proximity}, {\em cocircularity}, {\em smoothness} and {\em mass} have been taken into account is shown in blue. Note that the thick lines at $\log{k}=-2$ correspond to the range of optimum $\alpha$ values found (a) in all cases (black) and (b) when only the {\em proximity}, {\em cocircularity} and {\em smoothness} criteria are taken into account (red).}\label{alpha_summary}
\end{figure}

Apart from the $k$ parameter, the $\alpha$ parameter
also plays a key-role for the successful detection of patterns. We find a strong, positive correlation with the SN ratio, but more importantly, a strong correlation with the $k$ parameter. In Fig.\ref{alpha_summary} we present the $\log{k}-\alpha$ curves for all the tests discussed in the current work and for different number of interaction criteria used, as indicated within the figure. The $k-\alpha$ relation depends on the number of interaction coefficients used, becoming less steep when we increase the number of meaningful coefficients.\\Although we have found no accurate way of proposing values for the optimum $\alpha$ parameter in the whole range of $k$, from Fig. \ref{alpha_summary} it is evident that the range of the optimum $\alpha$ values, corresponding to $\log{k}\approx-2.$ (the best compromise, as discussed previously), is roughly  $0.02 \pm 0.015$. However, even within this restricted range, the resulting pattern remains relatively sensitive to the exact value of $\alpha$, indicating the necessity for further investigation depending on the pattern sought.

\section{ Application of the method on real data}

\subsection{A Cosmological N-Body Simulation}
For the current application we use one of the high-resolution Mare-Nostrum simulations of \citet{Tikhonov2009}. Specifically, we use a flat $\Lambda$CDM model with parameters $h = 0.73$, $\Omega_m = 0.24$, $\Omega_{b} = 0.042$, a power-spectrum normalization $\sigma_8 = 0.75$ and slope $n = 0.95$. The computational box is of 64 $h^{-1}$ Mpc with initially $4096^3$ dark matter (DM) particles, while after applying the Zeldovich approximation the number of DM particles was reduced to 1024$^{3}$, which corresponds to a mass per particle of $1.6\times 10^{7} h^{-1} M_\odot$. The simulation was performed using the TREEPM parallel N-body code GADGET2 \citep{Springel2005}, while the DM haloes were identified using the Adaptive Mesh Investigations of Galaxy Assembly (AMIGA) Halo Finder \citep{Knollmann2009}. For more details we guide the reader to read the original paper.

We apply our algorithm on the DM halo catalogue which is centered on a dominant filamentary structure extending through-out the 64 $h^{-1}$ Mpc box, with various smaller filaments joining the main structure. The application of our algorithm on the halo data is necessarily blind, in the sense that we do not have an {\em a priori} pre-determined pattern that we wish to identify, but rather we can test whether the sub-sample of haloes detected as pattern elements complies with the expectation of our pre-determined {\em interaction coefficients}. This highlights the fact that our methodology, as it stands currently, serves not as an absolute structure finder, but rather as a useful tool for the quantification of structures, exploiting the optimum range of values of the main parameters, ie., $\alpha < 0.050$ and $\log(k) = -2.$, identified from our Monte Carlo-based application. 
 
\begin{figure*}[p]
    \centering
    \begin{subfigure}[b]{0.475\textwidth}
        \includegraphics[width=0.927\textwidth]{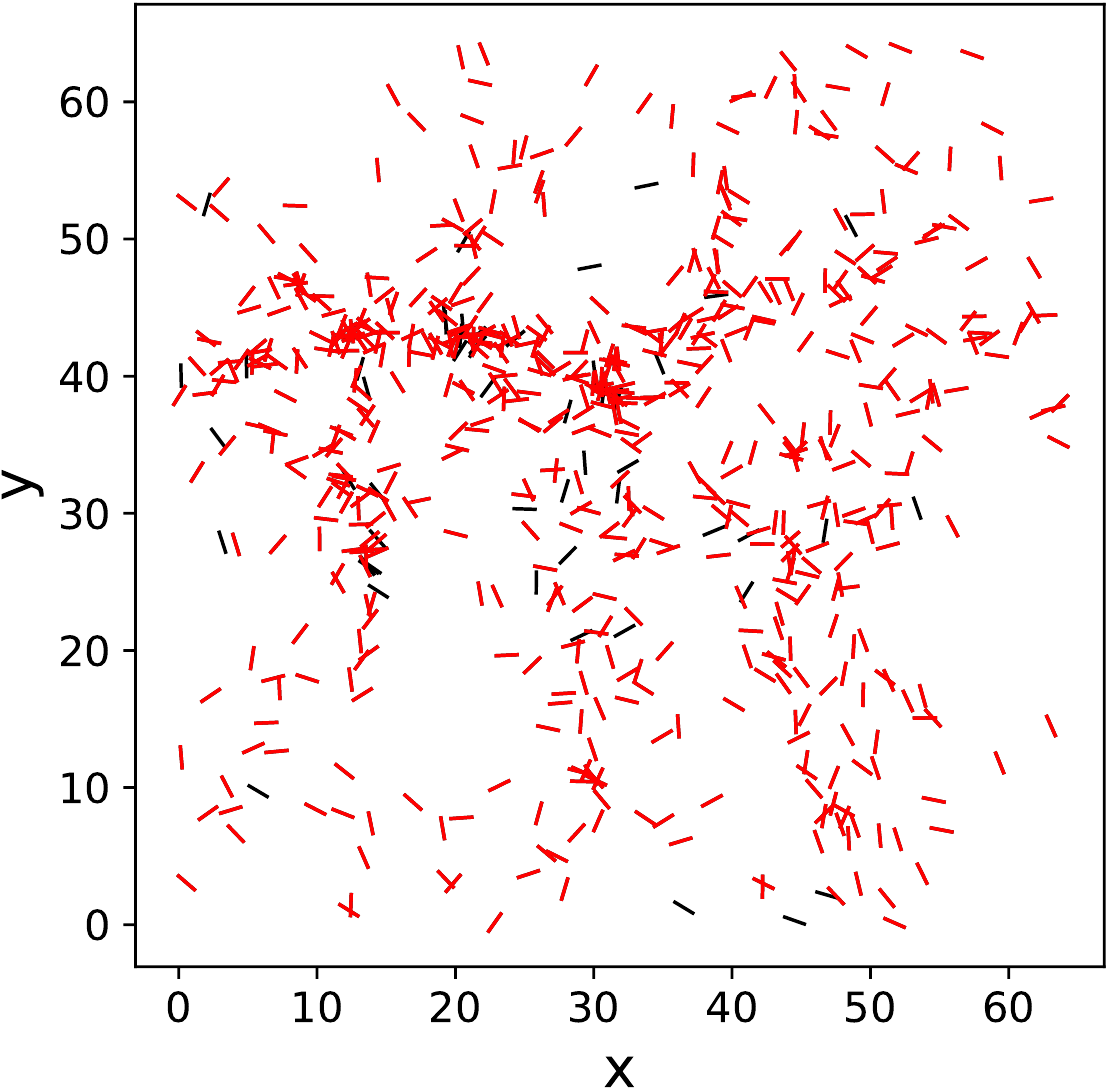}
    \end{subfigure}
    ~ 
    \begin{subfigure}[b]{0.475\textwidth}
        \includegraphics[width=0.927\textwidth]{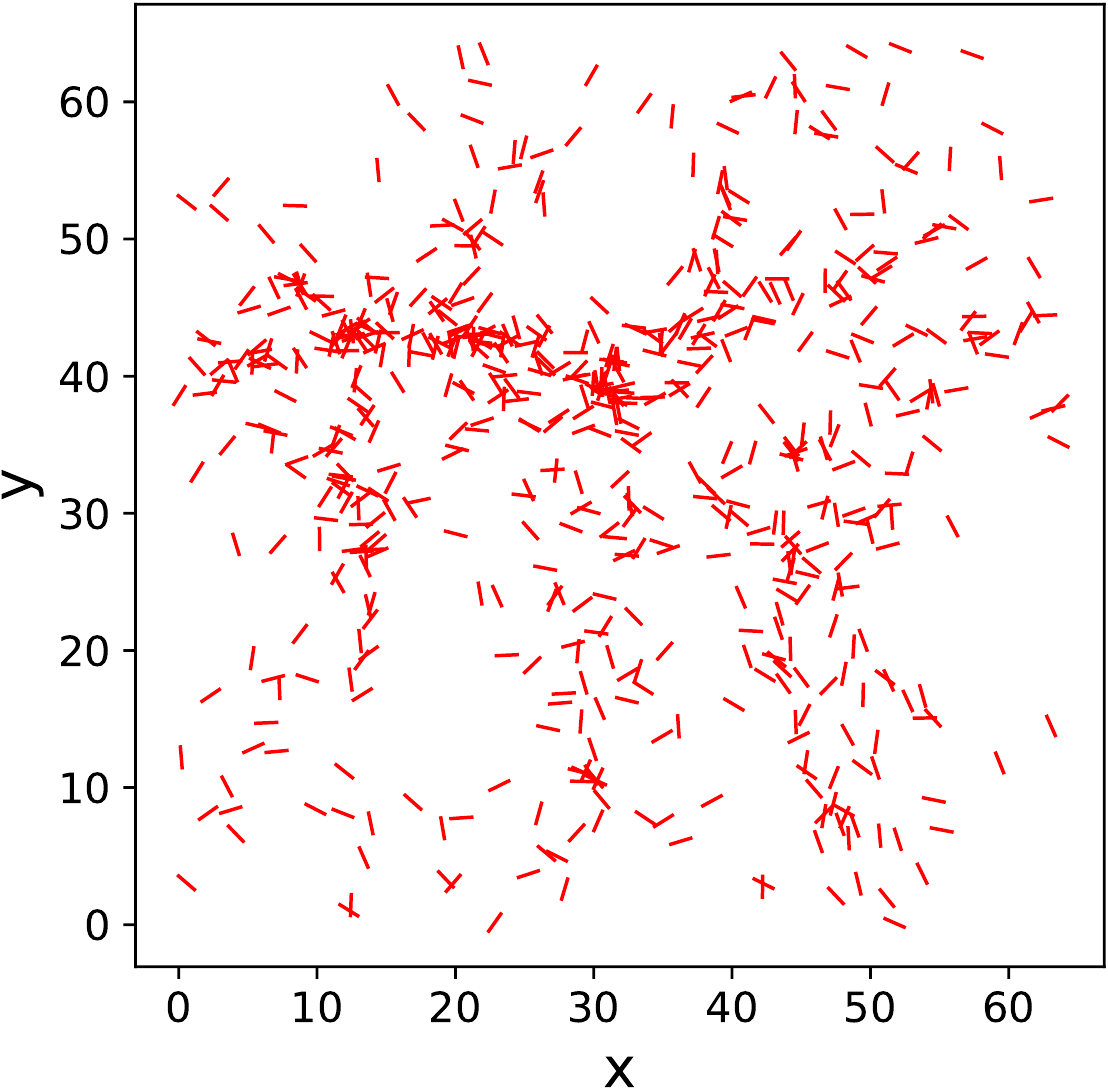}
    \end{subfigure}
    ~ 
    \begin{subfigure}[b]{0.475\textwidth}
        \includegraphics[width=0.927\textwidth]{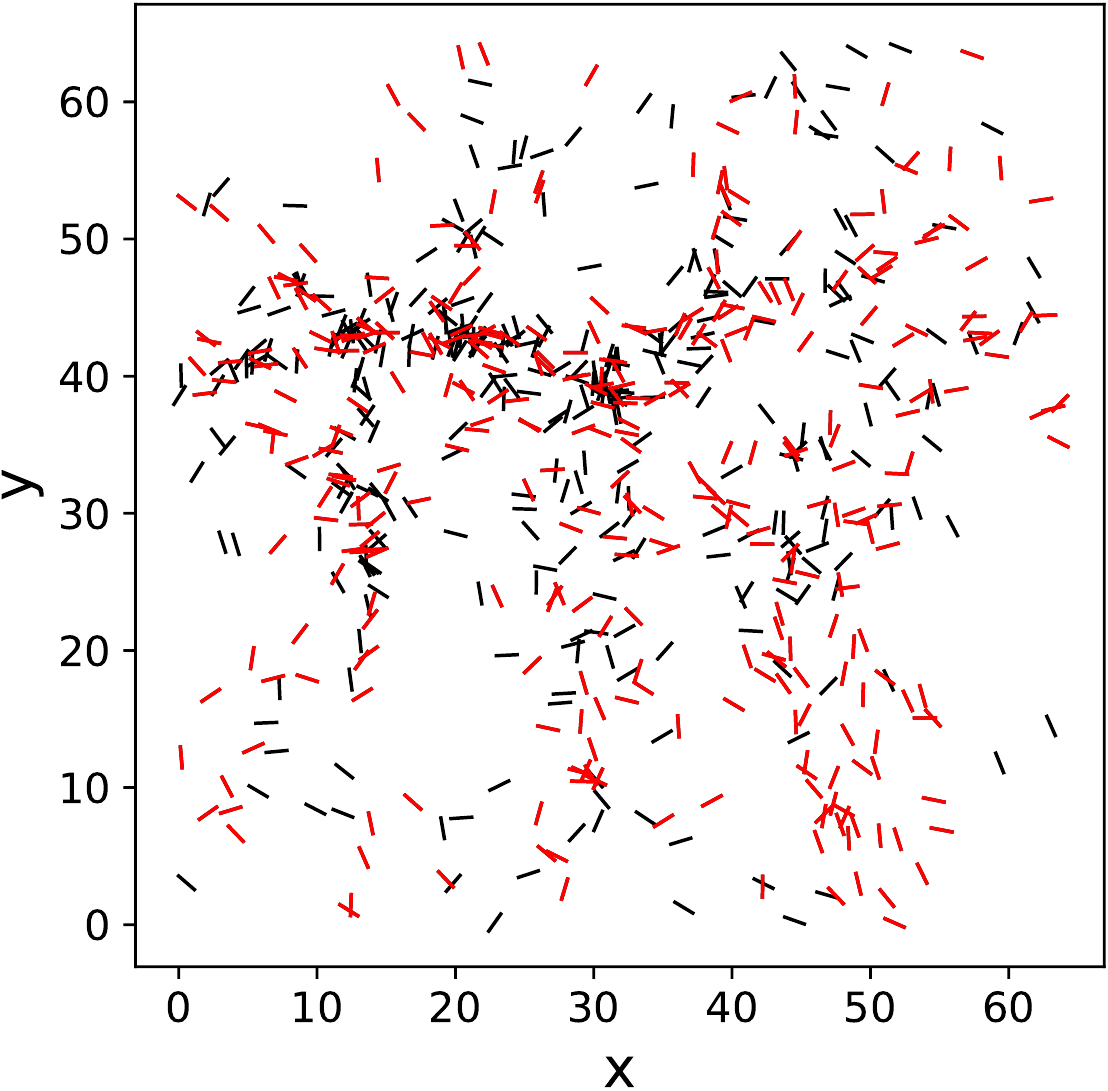}
    \end{subfigure}
    \begin{subfigure}[b]{0.475\textwidth}
        \includegraphics[width=0.927\textwidth]{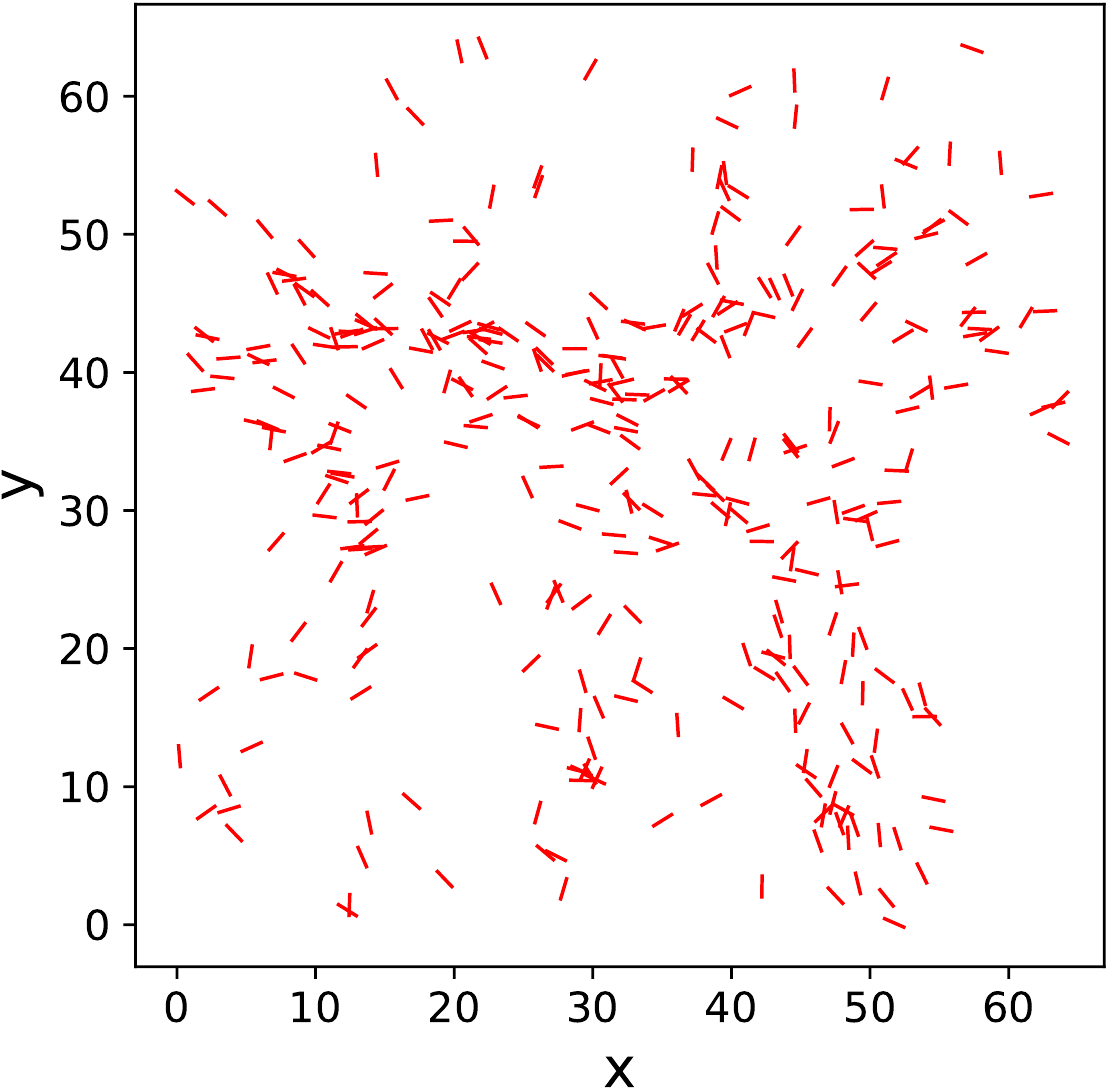}
    \end{subfigure}
    \begin{subfigure}[b]{0.475\textwidth}
        \includegraphics[width=0.927\textwidth]{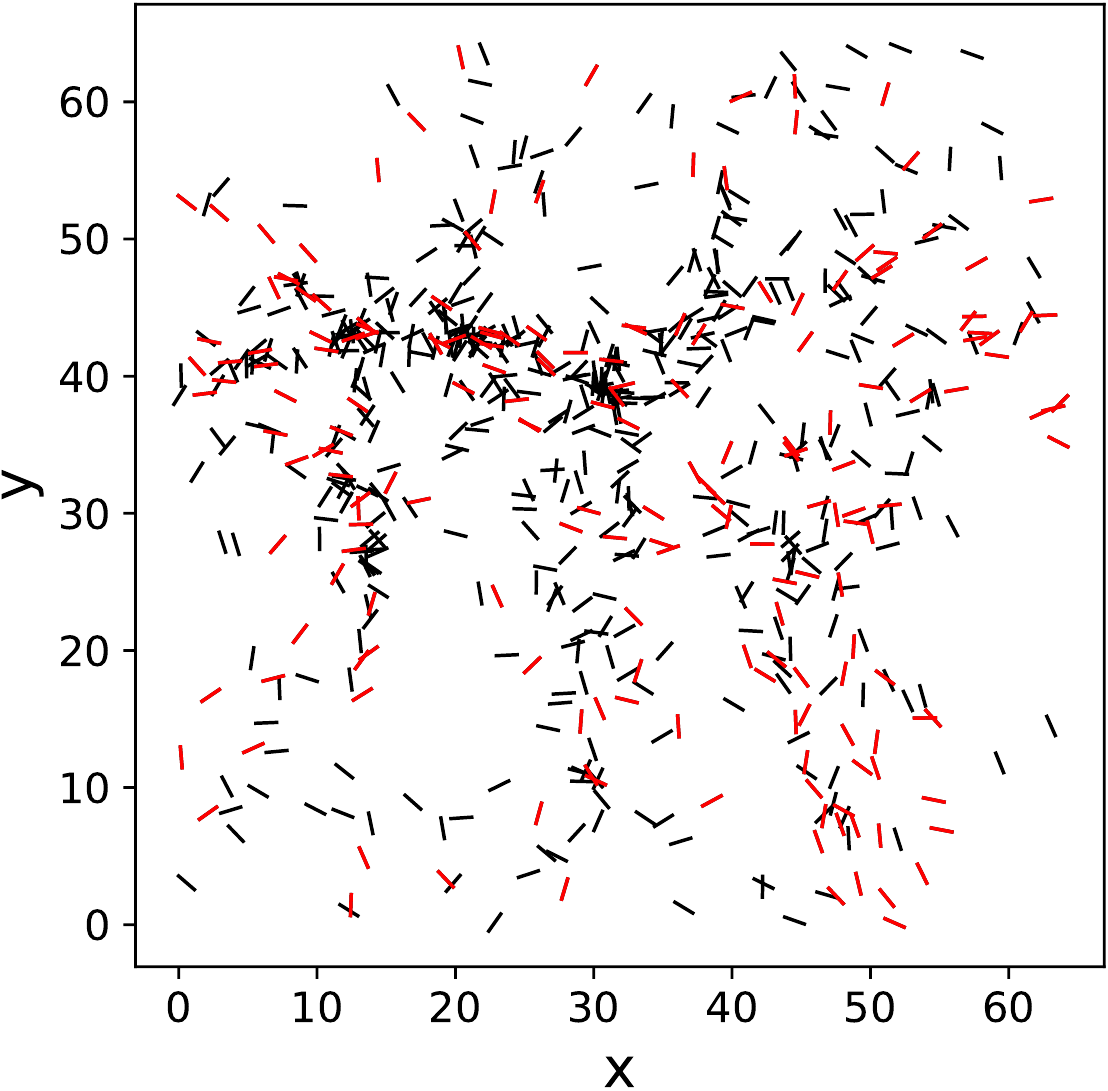}
    \end{subfigure}
    \begin{subfigure}[b]{0.475\textwidth}
        \includegraphics[width=0.927\textwidth]{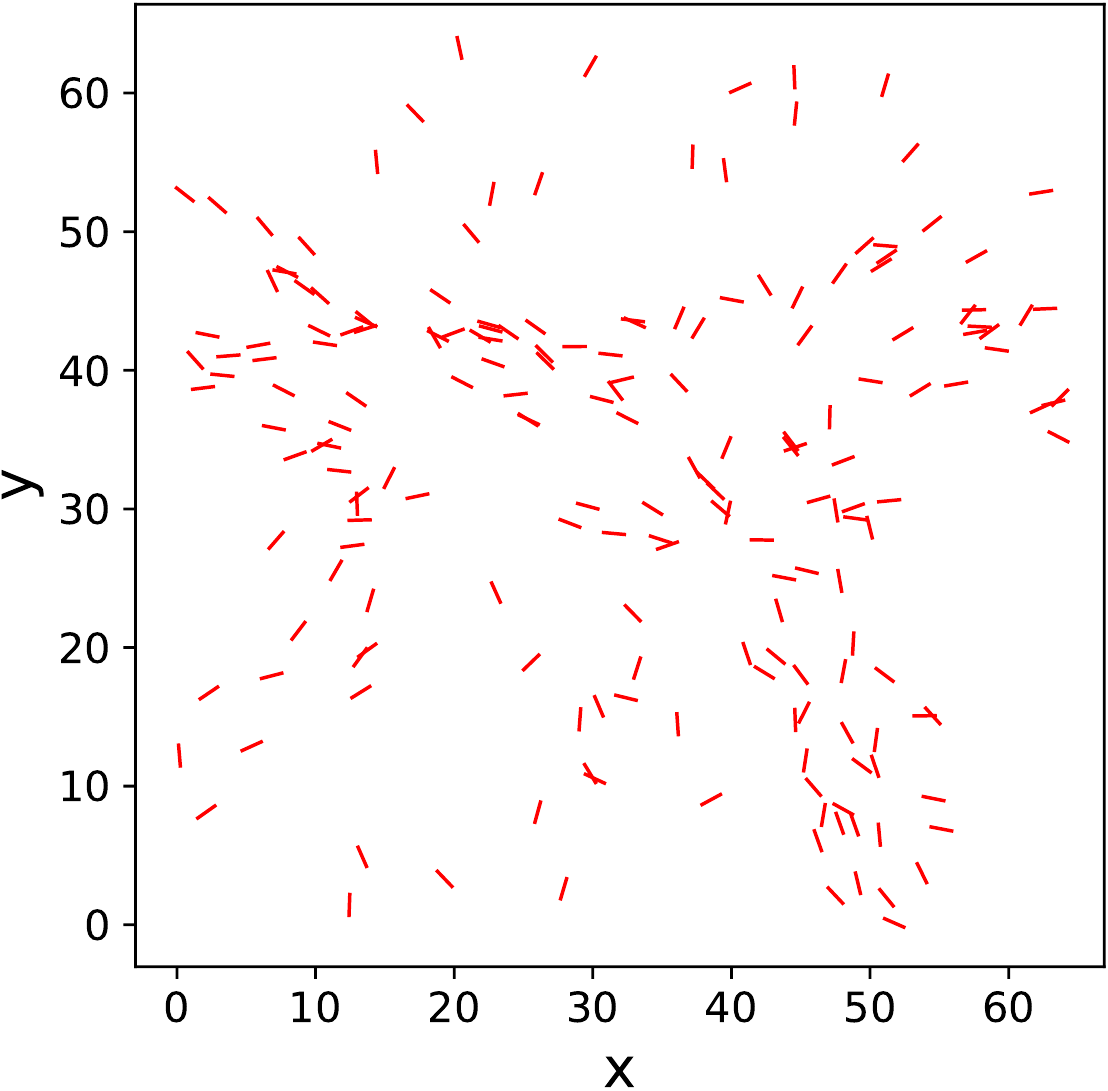}
    \end{subfigure}
    \caption{The results of applying the SA algorithm on the halo distribution of the {\em N}-body simulation snapshot for different values of $\alpha$, increasing from top to bottom. In the left-column images the pattern elements are shown in red while the rest of the halos (in black). In the right-columns we show only the pattern elements.}\label{fig:simulation_sequence}
\end{figure*}

 Some characteristic detected patterns, for sequential values of $\alpha$, are shown in Fig. \ref{fig:simulation_sequence}. For small values of $\alpha$ (e.g., $\alpha = 0.026$, top left) the algorithm tags the majority of elements as {\lq pattern-elements\rq}. This is consistent with the behavior of our algorithm in Section \ref{Application_Mock} for $\alpha$-values lower than the optimum. For slightly higher values of $\alpha$ the algorithm detects lower numbers of "pattern-elements", which, as can be seen in Fig. \ref{fig:simulation_sequence}, seem to have a greater degree of alignment-coherence. Indeed, based on our {\em interaction coefficients}, we should expect that the pattern to be detected should have a large degree of alignment coherence.
 
In order to quantify this visual impression, we devise a test to measure the degree of alignment coherence of the detected pattern as a function of the different $\alpha$-values. This consists of a measure of the local alignments, estimated within a sphere of some radius around each halo (here we use $R=3 h^{-1}$ Mpc), i.e. we estimate the misalignment angle, $\Delta\theta$, between the major axis of the central halo and that of each of its neighbors. In the case of random alignments, expected due to either projection effects or non-causally related structures, one expects a flat distribution between $0^o$ and $90^o$ of $\Delta\theta$, while in the case of coherent alignments, within the scale selected by the size of the sphere, one expects an anisotropic distribution with an excess of small $\Delta\theta$ values.

\begin{figure}
\includegraphics[scale = 0.4]{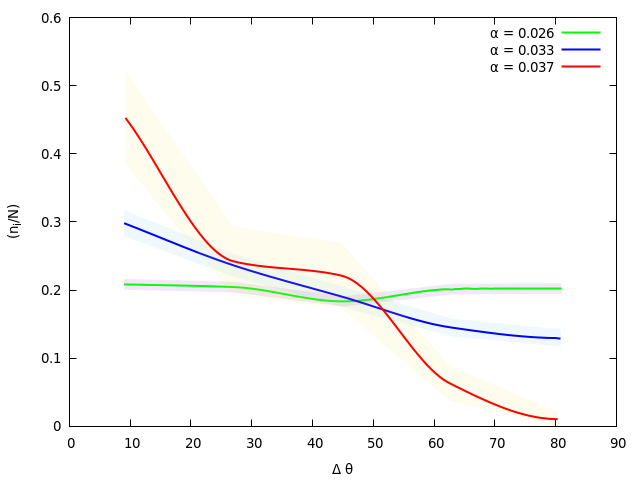}
\caption{The distributions of the small-scale alignments for different values of $\alpha$ parameter as denoted in the key.}\label{dtheta}
\end{figure}

 In Fig. \ref{dtheta} we present the $\Delta\theta$ distributions for three values of $\alpha = 0.026, 0.033, 0.037$, which are shown in green, blue and red respectively. The patterns detected have a much stronger alignment-coherence for higher values of $\alpha$, as can also be clearly seen in Fig. \ref{fig:simulation_sequence}. We wish to stress that although currently there is no objective way of defining an optimum or unique pattern, qualitative criteria as the one presented previously, i.e., that of the local alignment coherence, can be used depending on the specific interest of the algorithm user.

\subsection{Application on the lensing cluster Abell 2218}

We use the {\em HST} image of Abell 2218 (Credit: NASA, ESA, and Johan Richard - Caltech, USA), a rich galaxy cluster with thousands of galaxy members at a redshift of $z\simeq 0.17$ which acts as a powerful gravitational lens to magnify distant galaxies but also distorts them into long, thin arcs. Many tens of background lensed galaxies are present in the image, which covers a $\sim 3 \times 3$ arcmin$^2$ area with a resolution of 0.05 arcsec per pixel.

In order to detect the sources in the {\em HST} image we use the \texttt{PHOTUTILS} package\citep{Photutils}, which is an open source \texttt{PYTHON} affiliated package of {\em ASTROPY} that mainly provides tools for astronomical source detection and photometry. We select a Gaussian 2D Kernel of size $(x,y) = (5,5)$ and we also set the lower limit of connected pixels in each detected source to $n_{\rm pixels}=100$ and $contrast = 0.1$. The selection of these parameter values led to a satisfying compromise of noise-elimination and detection of small sources. The source detection process produces a full catalogue with the properties of each source, such as, the centroid Cartesian coordinates, orientation of the major axis, elongation, ellipticity, eccentricity, area, etc. Since after the source detection a significant level of background noise is still persistent, we put a source cut-off area $> 200$ pixels. Finally, during the algorithm application we set a lower limit on source-elongation $>1.7$. Such a cut-off is necessary in order to have robust determinations of the source major-axis orientation, since non-elongated sources have ill-defined orientations which will significantly affect the lens-candidate detection based on the criterion of {\em cocircularity}. 

\subsubsection{Pattern recognition using the 3 geometrical criteria}
\begin{figure*}
    \centering
    \includegraphics[width=0.85\textwidth]{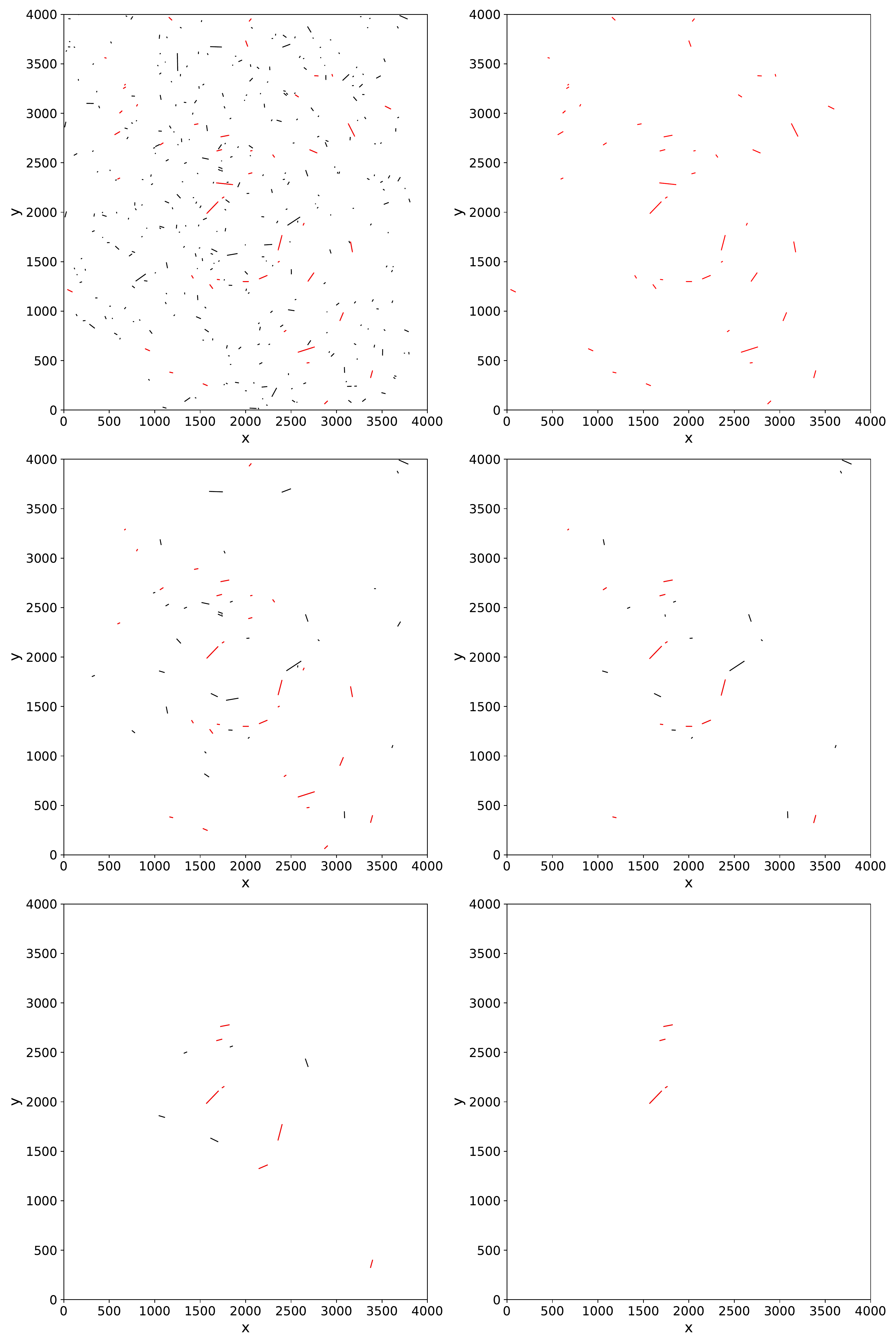}
    \caption{The sources detected by applying the Photutils package on the HST image of {\em Abell 1218}. The lens candidates selected via visual inspection are shown in red color on the image (top-left) and separately (top-right). Characteristic results of applying the SA algorithm on the image for different values of $\alpha$ are shown in the middle and bottom panels in black. The value of $\alpha$ increases from middle-left to bottom-right. Correctly detected lens-candidates are shown in red.}\label{fig:Abell_sequence}
\end{figure*}

We now apply our pattern recognition algorithm on the resulting source-catalogue with the aim of automatically detecting the lensed sources. To this end we first identify, via visual inspection, a relatively large number of lensed sources, some however with a significant level of uncertainty, and thus, we produce a sub-catalogue of sources that we tag as the {\lq lensed sources\rq}. These will be used to evaluate the results of the algorithm. All the sources detected using Photutils are shown in the top left-hand panel of Fig. \ref{fig:Abell_sequence}, where the 48 selected lens candidates are shown in red. For an easier comparison with the detected patterns, the lens candidates are also shown separately on the upper right-hand panel. The size of the markers is proportional to the  elongation of the source for a more meaningful symbolic representation. This is the case for all the following patterns presented in this paragraph. 

\begin{figure}
\includegraphics[width = 0.99\textwidth]{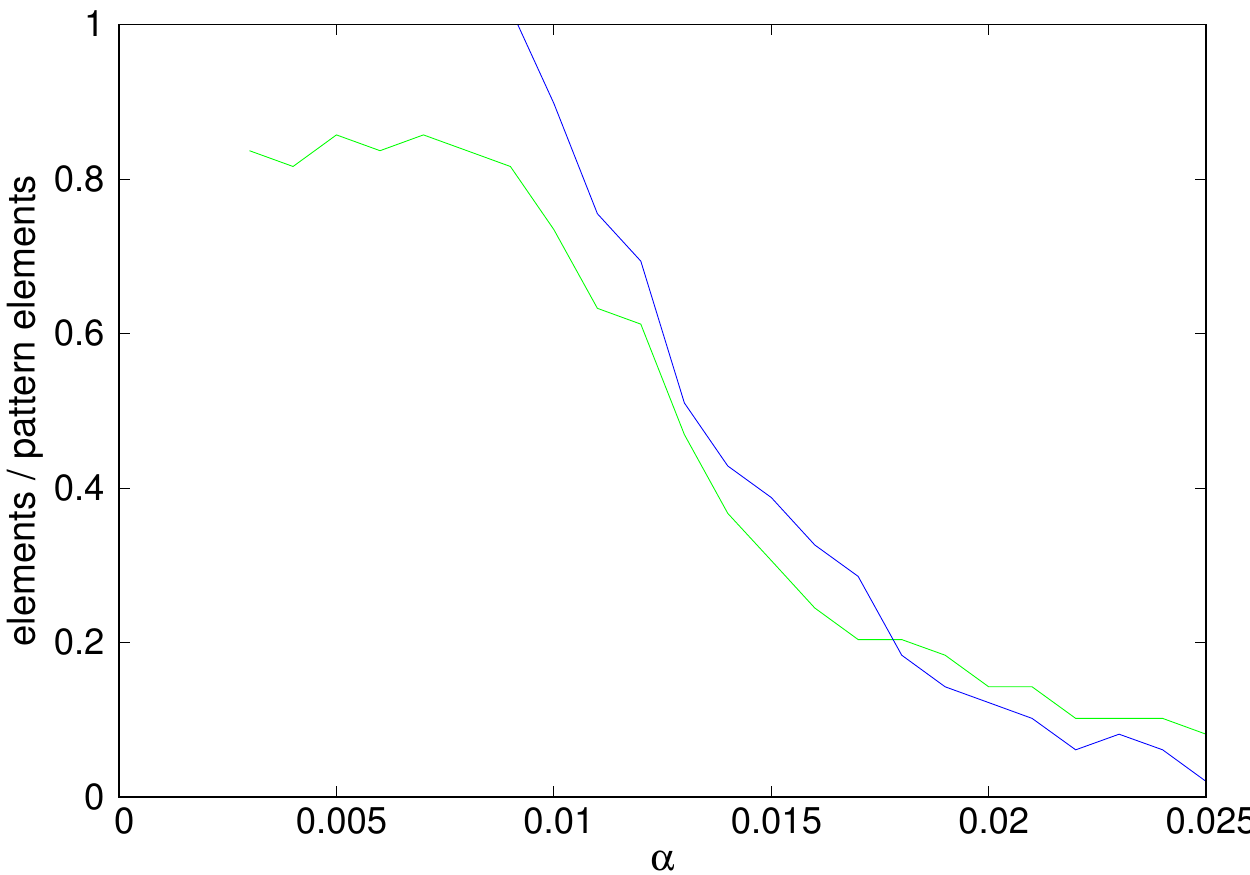}
\caption{Application on the {\em HST} Abell 2218 image. The detection quality of the algorithm as a function of $\alpha$ for $\log{k} = -2$. The green line corresponds to the image elements correctly tagged as {\lq pattern elements\rq} (or as {\lq lens-candidates\rq} for the specific application) and the blue line to the number of non-lensed sources {\lq falsely\rq} tagged as {\lq pattern elements\rq}. All values are normalized to the total number of lensed sources ($48$). }\label{Lens_results}
\end{figure}

We run our pattern recognition algorithm using the optimum parameter range, according to our Monte Carlo-based application; i.e. $\log(k) = -2$ and $\alpha < 0.050$.
For an overview of the evaluation of the current application, using as a reference the sub-catalogue containing the $48$ visually detected lens candidates, 

We present our results in Fig. \ref{Lens_results}, where  the fraction of the detected lens-candidates (in green) and the number of non-lensed sources, falsely characterized as such (in blue) and both normalized to the total number of lens candidates, are shown as a function of $\alpha$. Both drop with increasing $\alpha$, while for $\alpha \lesssim 0.01$, the detection of $~80$ per cent of the lens candidates comes with the cost of significant contamination by non-lensed sources. However, for $\alpha\gtrsim 0.017$ although a small fraction ($~ 10-20$ per cent) of the lens-candidates are detected by the algorithm, the fraction of {\lq false\rq} detections is lower than that of the true ones. Thus, for a secure detection of true lens candidates, with a relatively low contamination but also with significant loss of completeness, a high value of $\alpha$ is appropriate. On the other hand, if completeness is important and a significant level of contamination can be afforded, a low $\alpha$-value should be selected.
 
In order to visualize the outcome of the application of our algorithm, some characteristic cases of pattern-detections, for sequential values of $\alpha$, are shown in Fig. \ref{fig:Abell_sequence} (middle bottom panels). The detected candidates are shown in red, while "false" detections are shown in black. We mention that for $\alpha = 0.012$ (middle left-hand panel) the algorithm detects 30 lens-candidates out of 48 of our sub-catalogue while it also detects falsely, as "pattern-elements", $34$ other sources. For $\alpha = 0.026$ (bottom right-hand panel), there are four lens candidates detected with no contamination by {\lq false\rq} detections. It is evident that the detection of the majority of the lens-candidates comes with the cost of significant level of contamination. On the other hand, for higher values of $\alpha\sim 0.02$ a small number of sources are detected, but the level of contamination drops and is finally eliminated. 

One would have expected, from our relevant Monte Carlo analysis, under the assumption of a spherical and smooth potential, that the specific geometric criteria used would have provided satisfying results, which however is not the case. This poor performance should be probably attributed to the fact that {\em Abell 2218} is a substructured cluster\citep{Neumann1999} and thus the lensed sources are not expected to strictly comply with the cocircularity criterion. Furthermore, the possible inherent cluster galaxy alignments, due for example to the anisotropic cluster formation process, imply that part of the detected elements, although not arcs, could be correctly selected according to the specific interaction coefficients. Thus, our  definition of {\lq false\rq} elements may lead to underestimating the success of the algorithm.

\subsubsection{Pattern recognition using the elongation criterion}
Regarding the current application, we have verified that the imposing a lower-limit on elongation has a significant effect on the success of the algorithm. This can be explained in two ways. First, as we mentioned previously, the orientation of sources with low elongation carries large uncertainties and, since the definition of two interaction coefficients, {\em cocircularity} and {\em smoothness}, is based on the orientation of the image-elements, this is bound to affect severely the success of the algorithm. Secondly, lens candidates are in most cases quite elongated, and thus, putting a moderate limit on elongation serves as a first {\lq filter\rq} for the {\lq noise\rq} in the image.

A solution to such issues for the case of such a targeted application would be the use of an extra interaction coefficient, highlighting the specific physical properties of the structure of interest. Since, as discussed previously, elongation is such a property for the case of lensing phenomena, we perform an application of the algorithm inserting an extra coefficient, the {\em elongation coefficient}, which is defined, analogously to the {\em mass coefficient}, Section \ref{mass_definition}, as

\begin{ceqn}
\begin{align}
c_{i,j}^{\rm elong} = \frac{\rm elong_{i} \times elong_{j}}{\rm elong_{max}^{2}}\;\;,
 \label{elong}
\end{align}
\end{ceqn}
where ${\rm elong_{i}, elong_{j}}$ are the elongation of the members of the element-pair $({i,j})$ and ${\rm elong_{max}}$ is the elongation of the most elongated element in the image, used to normalize the coefficient to the range $[0,1]$.  

\begin{figure}
\includegraphics[width = 0.99\textwidth]{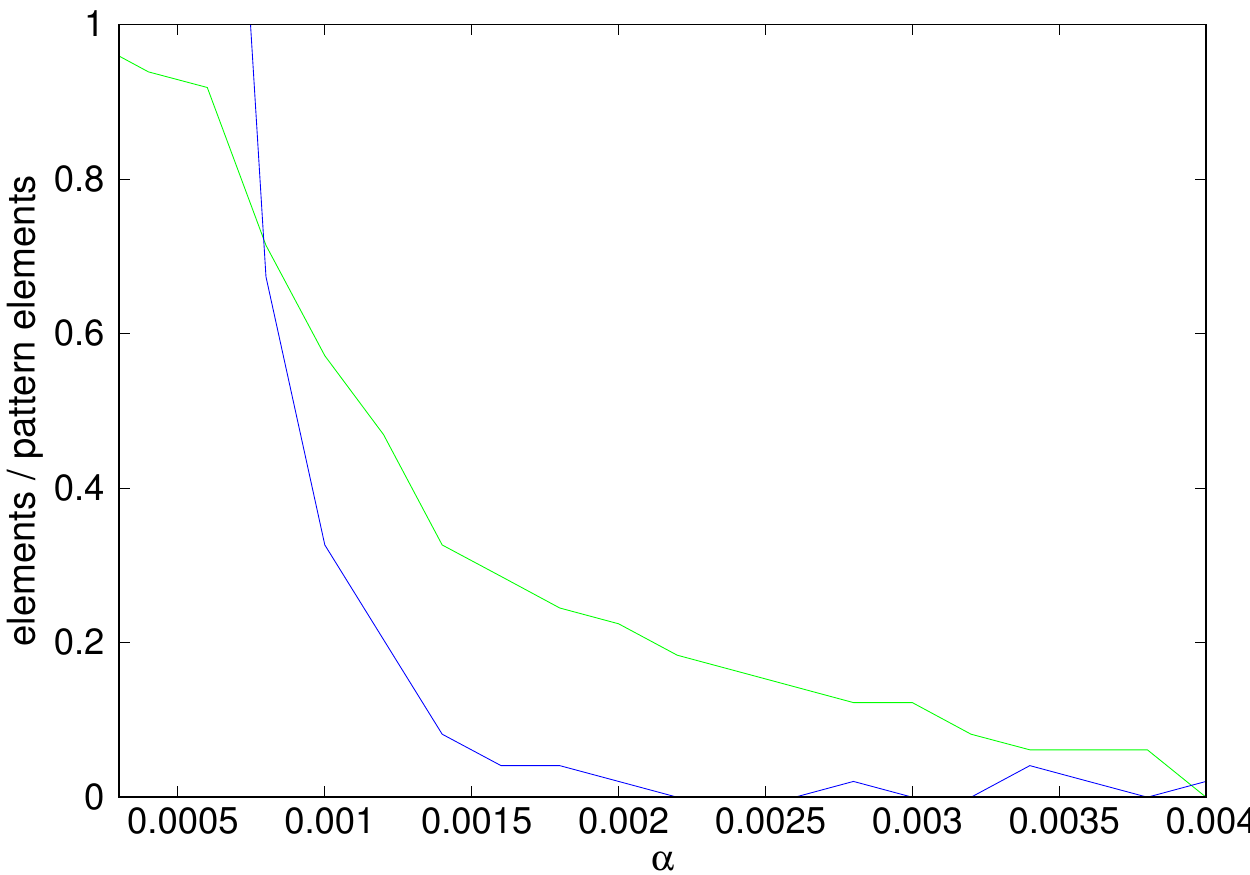}
\caption{The detection results of the algorithm as a function of $\alpha$ for $\log{k} = -2$, using additionally the {\em elongation coefficient}. The green line corresponds to the image elements correctly tagged as "pattern elements" (or as "lens-candidates" for the specific application) and the blue line to the number of non-lensed sources "falsely" tagged as "pattern elements". All values are normalized to the total number of lensed sources in our sub-catalogue ($48$). }\label{Lens_results_elong}
\end{figure}

The overall evaluation of the algorithm in the case of using additionally the {\em elongation coefficient} is presented in Fig. \ref{Lens_results_elong}. Evidently, the algorithm is significantly more successful in detecting lens candidates and for values of $\alpha \sim (0.013 - 0.02)$ the algorithm recovers a quite important fraction ($\sim 30-20$ per cent) of the lensed-source candidates with very low or zero contamination.
The optimum values of $\alpha$ drop, as expected when additional coefficients are used. In this case, the $\alpha$-values are of an order of magnitude lower than in the previous case, which probably implies that the {\em elongation coefficient} dominates the rest of the coefficients in this application. Note again that the high elongation of some cluster-galaxies imply that part of the detected elements, although not arcs, could be justifiably identified by the algorithm according to the new elongation interaction coefficient.

\begin{figure*}
    \centering
    \includegraphics[width=0.85\textwidth]{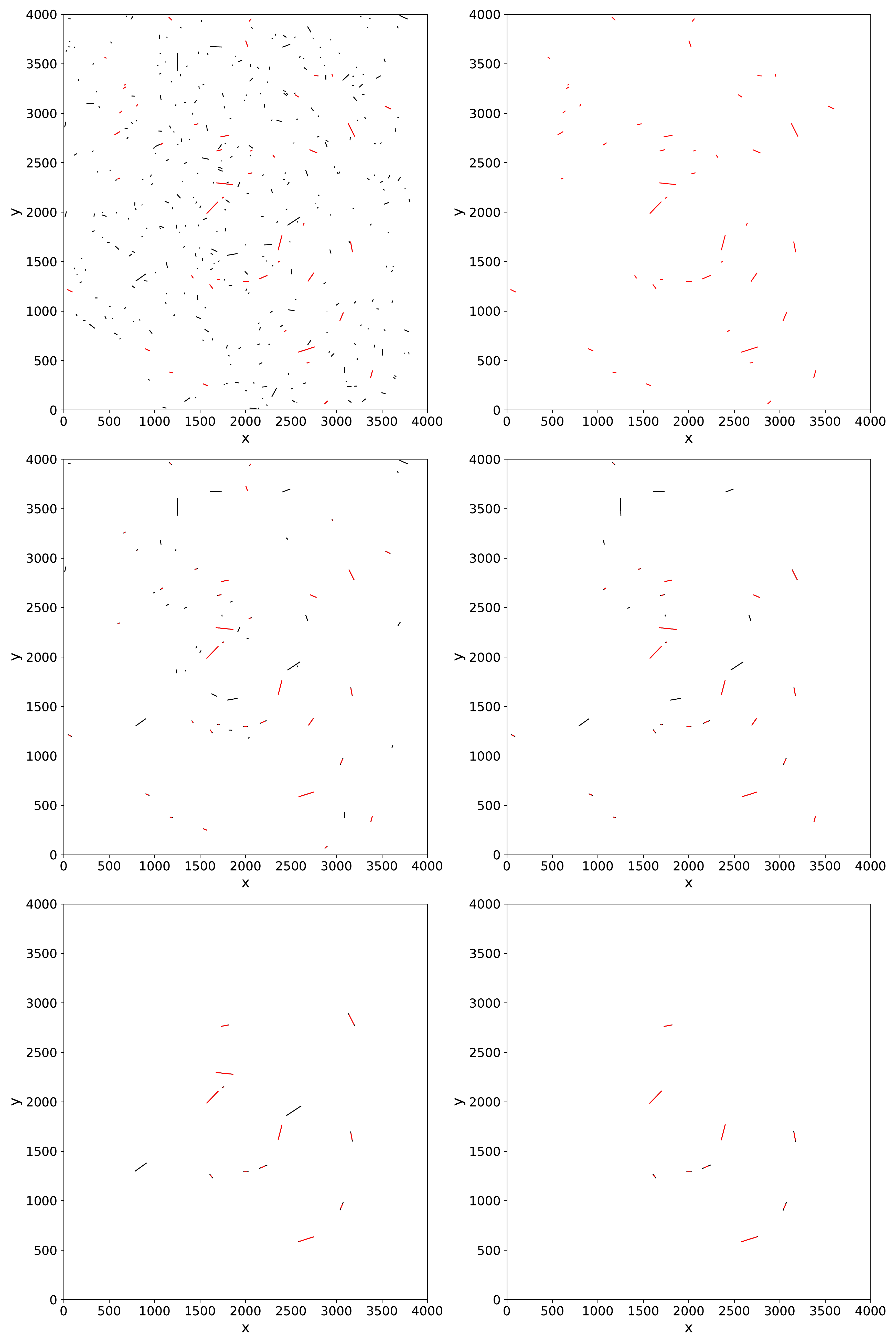}
    \caption{Similar as in Fig.\ref{fig:Abell_sequence} but for the case of using an additional interaction coefficient, that of the element elongation.
}\label{fig:Abell_sequence_elong}
\end{figure*}

The improvement is evident also in Fig. \ref{fig:Abell_sequence_elong} where we present a sequence of detected patterns for different values of $\alpha$. The detected candidates are shown in red, while the {\lq false\rq} detections in black. Indicatively, we mention that for $\alpha = 0.0008$ (middle left-hand panel), the algorithm detects 35 lens candidates and 33 non-lensed sources ({\lq false\rq}), while for $\alpha = 0.0022$ 9 lens-candidates are detected with zero contamination.

This is an indicative example of how this methodology could be adjusted and applied for the detection of patterns with specific a priori expected physical properties.

\section{Conclusions}
We applied a methodology to detect or quantify patterns on 2D images based on the figure ground discrimination techniques of \citet{HH1993}. We have constructed and used mock images that include: (a) circular patterns embedded in a background of noise elements with random positions and orientations and with number densities similar to that of massive DM halos in cosmological {\em N}-body simulations, (b) circular patterns, resembling the effect of strong gravitational lensing in the background of 419 SDSS galaxies (with $m_r<17.77$) of cluster {\em Abell 1656}. We have identified the optimum values of the free parameters of our algorithm, i.e. $\log k$ and $\alpha$, for the wide range of images discussed above. We have also successfully applied our algorithm to the distribution of haloes of a cosmological N-body simulation snap-shot, which is dominated by a filament-like structure, and identified patterns with the expected (based on the specific {\em interaction coefficients} used) property, i.e. strong alignment coherence. Finally, we have applied the algorithm on the optical image of lensing cluster {\em Abell 2218} and evaluated the results using as a reference the catalogue of lens candidates selected via visual inspection. 

To sum up, our main conclusions are as follows:
\begin{itemize}
\item The algorithm is quite successful in detecting the mock patterns tested, taking into account only the simplest geometrical interaction criteria of {\em proximity} and {\em cocircularity}.
\item Adding further the criterion of {\em smoothness} helps drastically in more realistic situations where the pair-wise alignment of the elements is not precise. 
\item Adding also non geometrical criteria, such as a criterion for mass or luminosity, can improve the results of the algorithm, depending on the physical properties of the sought patterns. 
\item The optimum value of the $\alpha$ parameter has an increasing tendency with the SN ratio, while it also depends on the value of $k$, used in the definition of the {\em cocircularity coefficient}.
\item A good compromise for the value of the $k$ parameter is $\log{k}\approx -2$, independently of the density of the image and the SN ratio. The respective range of the optimum $\alpha$ parameter is $\sim 0.02\pm0.015$. This range is expected to shrink and to shift towards lower values if more meaningful interaction coefficients are used in the algorithm.
\item The quantification of the interactions via the definition of different coefficients, renders this method versatile and suitable for the detection or quantification of patterns in a variety of problems, such as detecting coherent structures or the effects of gravitational lensing in cluster images. Indeed, a blind application on the halo catalogue of an N-body simulation snap-shot, has revealed that we can successfully extract the counterpart of cosmic patterns with the expected (based on the specific {\em interaction coefficients} used) properties. Finally, we also applied the algorithm on the optical {\em HST} image of cluster {\em Abell 2218}. The application of the basic algorithm did not provide satisfying results since the cluster potential is substructured, and thus the lensed sources do not strictly comply to the cocircularity criterion. However, the use of an extra {\em elongation coefficient}, a possibility highlighting the versatility of the algorithm, leads to improved results indicating that such a methodology could be used as a rough {\lq filter\rq} in order to extract lensed source candidates from a large number of images for which individual visual inspection could not be afforded.
\end{itemize}

\section*{Acknowledgements}

This research is co-financed by Greece and the European Union (European Social Fund- ESF) through the Operational Programme "Human Resources Development, Education and Lifelong Learning" in the context of the project "Strengthening Human Resources Research Potential via Doctorate Research" (MIS-5000432), implemented by the State Scholarships Foundation (IKY).

 We acknowledge Davide de Martin \& James Long (ESA/Hubble) for use of the {\em HST} image of A2218, as well as Prof. Gustavo Yepes for providing access to the Cosmological N-body Simulation data used in the current work.  We also wish to thank Dr. Angel-Ruiz Camu{\~n}as for his help in using the \texttt{PHOTUTILS} package.

This research made use of \texttt{PHOTUTILS}, an \texttt{ASTROPY} package for detection and photometry of astronomical sources.

Finally, we would like to thank the anonymous referee for suggestions that improved significantly the work presented in our manuscript.







\bsp	
\label{lastpage}

\begin{thebibliography}{}
\makeatletter
\relax
\def\mn@urlcharsother{\let\do\@makeother \do\$\do\&\do\#\do\^\do\_\do\%\do\~}
\def\mn@doi{\begingroup\mn@urlcharsother \@ifnextchar [ {\mn@doi@}
  {\mn@doi@[]}}
\def\mn@doi@[#1]#2{\def\@tempa{#1}\ifx\@tempa\@empty \href
  {http://dx.doi.org/#2} {doi:#2}\else \href {http://dx.doi.org/#2} {#1}\fi
  \endgroup}
\def\mn@eprint#1#2{\mn@eprint@#1:#2::\@nil}
\def\mn@eprint@arXiv#1{\href {http://arxiv.org/abs/#1} {{\tt arXiv:#1}}}
\def\mn@eprint@dblp#1{\href {http://dblp.uni-trier.de/rec/bibtex/#1.xml}
  {dblp:#1}}
\def\mn@eprint@#1:#2:#3:#4\@nil{\def\@tempa {#1}\def\@tempb {#2}\def\@tempc
  {#3}\ifx \@tempc \@empty \let \@tempc \@tempb \let \@tempb \@tempa \fi \ifx
  \@tempb \@empty \def\@tempb {arXiv}\fi \@ifundefined
  {mn@eprint@\@tempb}{\@tempb:\@tempc}{\expandafter \expandafter \csname
  mn@eprint@\@tempb\endcsname \expandafter{\@tempc}}}

\bibitem[\protect\citeauthoryear{{Ahn} et~al.,}{{Ahn} et~al.}{2014}]{Ahn2014}
{Ahn} C.~P.,  et~al., 2014, \mn@doi [\apjs] {10.1088/0067-0049/211/2/17}, \href
  {http://adsabs.harvard.edu/abs/2014ApJS..211...17A} {211, 17}

\bibitem[\protect\citeauthoryear{{Alimi}, {F{\"u}zfa}, {Boucher}, {Rasera},
  {Courtin}  \& {Corasaniti}}{{Alimi} et~al.}{2010}]{Alimi2010}
{Alimi} J.-M.,  {F{\"u}zfa} A.,  {Boucher} V.,  {Rasera} Y.,  {Courtin} J.,
  {Corasaniti} P.-S.,  2010, \mn@doi [MNRAS]
  {10.1111/j.1365-2966.2009.15712.x}, \href
  {http://adsabs.harvard.edu/abs/2010MNRAS.401..775A} {401, 775}

\bibitem[\protect\citeauthoryear{{Andr{\'e}}, {K{\"o}nyves}, {Arzoumanian},
  {Palmeirim}  \& {Peretto}}{{Andr{\'e}} et~al.}{2013}]{Andre2013}
{Andr{\'e}} P.,  {K{\"o}nyves} V.,  {Arzoumanian} D.,  {Palmeirim} P.,
  {Peretto} N.,  2013, in {Kawabe} R.,  {Kuno} N.,   {Yamamoto} S.,  eds,
  Astronomical Society of the Pacific Conference Series Vol. 476, New Trends in
  Radio Astronomy in the ALMA Era: The 30th Anniversary of Nobeyama Radio
  Observatory. p.~95

\bibitem[\protect\citeauthoryear{{Bartelmann}}{{Bartelmann}}{2010}]{Bartelmann2010}
{Bartelmann} M.,  2010, \mn@doi [Classical and Quantum Gravity]
  {10.1088/0264-9381/27/23/233001}, \href
  {http://adsabs.harvard.edu/abs/2010CQGra..27w3001B} {27, 233001}

\bibitem[\protect\citeauthoryear{{Binggeli}}{{Binggeli}}{1982}]{Binggeli1982}
{Binggeli} B.,  1982, \aap, \href
  {https://ui.adsabs.harvard.edu/abs/1982A&A...107..338B} {107, 338}

\bibitem[\protect\citeauthoryear{{Bond}, {Kofman}  \& {Pogosyan}}{{Bond}
  et~al.}{1996}]{Bond1996}
{Bond} J.~R.,  {Kofman} L.,   {Pogosyan} D.,  1996, \mn@doi [Nature]
  {10.1038/380603a0}, \href {http://adsabs.harvard.edu/abs/1996Natur.380..603B}
  {380, 603}

\bibitem[\protect\citeauthoryear{Bradley et~al.,}{Bradley
  et~al.}{2019}]{Photutils}
Bradley L.,  et~al., 2019, astropy/photutils: v0.7,
  \mn@doi{10.5281/zenodo.3368647}, \url
  {https://doi.org/10.5281/zenodo.3368647}

\bibitem[\protect\citeauthoryear{{Chira}, {Plionis}  \& {Corasaniti}}{{Chira}
  et~al.}{2018}]{Chira2018}
{Chira} M.,  {Plionis} M.,   {Corasaniti} P.-S.,  2018, \mn@doi [A\&A]
  {10.1051/0004-6361/201731440}, \href
  {http://adsabs.harvard.edu/abs/2018A%26A...616A.137C} {616, A137}

\bibitem[\protect\citeauthoryear{{Courtin}, {Rasera}, {Alimi}, {Corasaniti},
  {Boucher}  \& {F{\"u}zfa}}{{Courtin} et~al.}{2011}]{Courtin2011}
{Courtin} J.,  {Rasera} Y.,  {Alimi} J.-M.,  {Corasaniti} P.-S.,  {Boucher} V.,
    {F{\"u}zfa} A.,  2011, \mn@doi [MNRAS] {10.1111/j.1365-2966.2010.17573.x},
  \href {http://adsabs.harvard.edu/abs/2011MNRAS.410.1911C} {410, 1911}

\bibitem[\protect\citeauthoryear{{Eckrot}, {Geldhauser}  \& {Jurczyk}}{{Eckrot}
  et~al.}{2017}]{Eckrot2017}
{Eckrot} A.,  {Geldhauser} C.,   {Jurczyk} J.,  2017, preprint, \href
  {http://adsabs.harvard.edu/abs/2017arXiv170401049E} {} (\mn@eprint {arXiv}
  {1704.01049})

\bibitem[\protect\citeauthoryear{{Faltenbacher}, {Gottl{\"o}ber}, {Kerscher}
  \& {M{\"u}ller}}{{Faltenbacher} et~al.}{2002}]{Faltenbacher2002}
{Faltenbacher} A.,  {Gottl{\"o}ber} S.,  {Kerscher} M.,   {M{\"u}ller} V.,
  2002, \mn@doi [\aap] {10.1051/0004-6361:20021263}, \href
  {https://ui.adsabs.harvard.edu/abs/2002A&A...395....1F} {395, 1}

\bibitem[\protect\citeauthoryear{{Ganeshaiah Veena}, {Cautun}, {Tempel}, {van
  de Weygaert}  \& {Frenk}}{{Ganeshaiah Veena} et~al.}{2019}]{Veena2019}
{Ganeshaiah Veena} P.,  {Cautun} M.,  {Tempel} E.,  {van de Weygaert} R.,
  {Frenk} C.~S.,  2019, arXiv e-prints, \href
  {https://ui.adsabs.harvard.edu/abs/2019arXiv190306716G} {p. arXiv:1903.06716}

\bibitem[\protect\citeauthoryear{Geman \& Geman}{Geman \&
  Geman}{1984}]{Geman1984}
Geman S.,  Geman D.,  1984, \mn@doi [IEEE Transactions on Pattern Analysis and
  Machine Intelligence] {10.1109/TPAMI.1984.4767596}, PAMI-6, 721

\bibitem[\protect\citeauthoryear{{Habib}, {Vernin}, {Benkhaldoun}  \&
  {Lanteri}}{{Habib} et~al.}{2006}]{Habib2006}
{Habib} A.,  {Vernin} J.,  {Benkhaldoun} Z.,   {Lanteri} H.,  2006, \mn@doi
  [MNRAS] {10.1111/j.1365-2966.2006.10235.x}, \href
  {http://adsabs.harvard.edu/abs/2006MNRAS.368.1456H} {368, 1456}

\bibitem[\protect\citeauthoryear{Herault \& Horaud}{Herault \&
  Horaud}{1993}]{HH1993}
Herault L.,  Horaud R.,  1993, \mn@doi [IEEE Transactions on Pattern Analysis
  and Machine Intelligence] {10.1109/34.232076}, 15, 899

\bibitem[\protect\citeauthoryear{{Joachimi} et~al.,}{{Joachimi}
  et~al.}{2015}]{Joachimi2015}
{Joachimi} B.,  et~al., 2015, \mn@doi [\ssr] {10.1007/s11214-015-0177-4}, \href
  {https://ui.adsabs.harvard.edu/abs/2015SSRv..193....1J} {193, 1}

\bibitem[\protect\citeauthoryear{{Kasun} \& {Evrard}}{{Kasun} \&
  {Evrard}}{2005}]{Kasun2005}
{Kasun} S.~F.,  {Evrard} A.~E.,  2005, \mn@doi [\apj] {10.1086/430811}, \href
  {https://ui.adsabs.harvard.edu/abs/2005ApJ...629..781K} {629, 781}

\bibitem[\protect\citeauthoryear{{Keelan}, {Chung}  \& {Hague}}{{Keelan}
  et~al.}{2018}]{Keelan2018}
{Keelan} J.,  {Chung} E.~M.~L.,   {Hague} J.~P.,  2018, preprint, \href
  {http://adsabs.harvard.edu/abs/2018arXiv180711513K} {} (\mn@eprint {arXiv}
  {1807.11513})

\bibitem[\protect\citeauthoryear{{Kirkpatrick}, {Gelatt}  \&
  {Vecchi}}{{Kirkpatrick} et~al.}{1983}]{Kirk1983}
{Kirkpatrick} S.,  {Gelatt} C.~D.,   {Vecchi} M.~P.,  1983, \mn@doi [Science]
  {10.1126/science.220.4598.671}, \href
  {http://adsabs.harvard.edu/abs/1983Sci...220..671K} {220, 671}

\bibitem[\protect\citeauthoryear{{Knollmann} \& {Knebe}}{{Knollmann} \&
  {Knebe}}{2009}]{Knollmann2009}
{Knollmann} S.~R.,  {Knebe} A.,  2009, \mn@doi [\apjs]
  {10.1088/0067-0049/182/2/608}, \href
  {https://ui.adsabs.harvard.edu/abs/2009ApJS..182..608K} {182, 608}

\bibitem[\protect\citeauthoryear{{Kochanek}, {Keeton}  \& {McLeod}}{{Kochanek}
  et~al.}{2001}]{Kochanek2001}
{Kochanek} C.~S.,  {Keeton} C.~R.,   {McLeod} B.~A.,  2001, \mn@doi [\apj]
  {10.1086/318350}, \href {http://adsabs.harvard.edu/abs/2001ApJ...547...50K}
  {547, 50}

\bibitem[\protect\citeauthoryear{{Kovalenko}, {Stoica}  \&
  {Emelyanov}}{{Kovalenko} et~al.}{2017}]{Kovalenko2017}
{Kovalenko} I.~D.,  {Stoica} R.~S.,   {Emelyanov} N.~V.,  2017, \mn@doi [MNRAS]
  {10.1093/mnras/stx1899}, \href
  {http://adsabs.harvard.edu/abs/2017MNRAS.471.4637K} {471, 4637}

\bibitem[\protect\citeauthoryear{{Libeskind} et~al.,}{{Libeskind}
  et~al.}{2018}]{Libeskind2018}
{Libeskind} N.~I.,  et~al., 2018, \mn@doi [\mnras] {10.1093/mnras/stx1976},
  \href {http://adsabs.harvard.edu/abs/2018MNRAS.473.1195L} {473, 1195}

\bibitem[\protect\citeauthoryear{{Neumann} \& {B{\"o}hringer}}{{Neumann} \&
  {B{\"o}hringer}}{1999}]{Neumann1999}
{Neumann} D.~M.,  {B{\"o}hringer} H.,  1999, \mn@doi [\apj] {10.1086/306812},
  \href {https://ui.adsabs.harvard.edu/abs/1999ApJ...512..630N} {512, 630}

\bibitem[\protect\citeauthoryear{{Pandya} et~al.,}{{Pandya}
  et~al.}{2019}]{Pandya2019}
{Pandya} V.,  et~al., 2019, arXiv e-prints, \href
  {https://ui.adsabs.harvard.edu/abs/2019arXiv190209559P} {p. arXiv:1902.09559}

\bibitem[\protect\citeauthoryear{Pham \& Karaboga}{Pham \&
  Karaboga}{2000}]{Pham}
Pham D.,  Karaboga D.,  2000, Intelligent optimisation techniques. Genetic
  algorithms, tabu search, simulated annealing and neural networks

\bibitem[\protect\citeauthoryear{{Plionis}}{{Plionis}}{1994}]{Plionis1994}
{Plionis} M.,  1994, \mn@doi [\apjs] {10.1086/192104}, \href
  {https://ui.adsabs.harvard.edu/abs/1994ApJS...95..401P} {95, 401}

\bibitem[\protect\citeauthoryear{{Rasera}, {Alimi}, {Courtin}, {Roy},
  {Corasaniti}, {F{\"u}zfa}  \& {Boucher}}{{Rasera} et~al.}{2010}]{Rasera2010}
{Rasera} Y.,  {Alimi} J.-M.,  {Courtin} J.,  {Roy} F.,  {Corasaniti} P.-S.,
  {F{\"u}zfa} A.,   {Boucher} V.,  2010, in {Alimi} J.-M.,  {Fu{\"o}zfa} A.,
  eds,  American Institute of Physics Conference Series Vol. 1241, American
  Institute of Physics Conference Series. pp 1134--1139 (\mn@eprint {arXiv}
  {1002.4950}), \mn@doi{10.1063/1.3462610}

\bibitem[\protect\citeauthoryear{{Refsdal} \& {Surdej}}{{Refsdal} \&
  {Surdej}}{1994}]{Refsdal1994}
{Refsdal} S.,  {Surdej} J.,  1994, \mn@doi [Reports on Progress in Physics]
  {10.1088/0034-4885/57/2/001}, \href
  {http://adsabs.harvard.edu/abs/1994RPPh...57..117R} {57, 117}

\bibitem[\protect\citeauthoryear{Sonmez}{Sonmez}{2007}]{Sonmez2007}
Sonmez F.~O.,  2007.

\bibitem[\protect\citeauthoryear{{Springel}}{{Springel}}{2005}]{Springel2005}
{Springel} V.,  2005, \mn@doi [\mnras] {10.1111/j.1365-2966.2005.09655.x},
  \href {https://ui.adsabs.harvard.edu/abs/2005MNRAS.364.1105S} {364, 1105}

\bibitem[\protect\citeauthoryear{Stoica, Gregori  \& Mateu}{Stoica
  et~al.}{2005}]{STOICA2005}
Stoica R.,  Gregori P.,   Mateu J.,  2005, \mn@doi [Stochastic Processes and
  their Applications] {https://doi.org/10.1016/j.spa.2005.06.007}, 115, 1860

\bibitem[\protect\citeauthoryear{{Stoica}, {Martinez}  \& {Saar}}{{Stoica}
  et~al.}{2008}]{Stoica2008}
{Stoica} R.~S.,  {Martinez} V.~J.,   {Saar} E.,  2008, arXiv e-prints, \href
  {https://ui.adsabs.harvard.edu/abs/2008arXiv0809.4358S} {p. arXiv:0809.4358}

\bibitem[\protect\citeauthoryear{Storvik}{Storvik}{1994}]{Storvik1994}
Storvik G.,  1994, \mn@doi [IEEE Transactions on Pattern Analysis and Machine
  Intelligence] {10.1109/34.329011}, 16, 976

\bibitem[\protect\citeauthoryear{{Tempel}, {Libeskind}, {Hoffman},
  {Liivam{\"a}gi}  \& {Tamm}}{{Tempel} et~al.}{2014a}]{Tempel2014b}
{Tempel} E.,  {Libeskind} N.~I.,  {Hoffman} Y.,  {Liivam{\"a}gi} L.~J.,
  {Tamm} A.,  2014a, \mn@doi [\mnras] {10.1093/mnrasl/slt130}, \href
  {https://ui.adsabs.harvard.edu/abs/2014MNRAS.437L..11T} {437, L11}

\bibitem[\protect\citeauthoryear{{Tempel}, {Stoica}, {Mart{\'\i}nez},
  {Liivam{\"a}gi}, {Castellan}  \& {Saar}}{{Tempel} et~al.}{2014b}]{Tempel2014}
{Tempel} E.,  {Stoica} R.~S.,  {Mart{\'\i}nez} V.~J.,  {Liivam{\"a}gi} L.~J.,
  {Castellan} G.,   {Saar} E.,  2014b, \mn@doi [\mnras]
  {10.1093/mnras/stt2454}, \href
  {https://ui.adsabs.harvard.edu/abs/2014MNRAS.438.3465T} {438, 3465}

\bibitem[\protect\citeauthoryear{{Tempel}, {Stoica}, {Kipper}  \&
  {Saar}}{{Tempel} et~al.}{2016}]{Tempel2016}
{Tempel} E.,  {Stoica} R.~S.,  {Kipper} R.,   {Saar} E.,  2016, \mn@doi
  [Astronomy and Computing] {10.1016/j.ascom.2016.03.004}, \href
  {https://ui.adsabs.harvard.edu/abs/2016A&C....16...17T} {16, 17}

\bibitem[\protect\citeauthoryear{{Tempel}, {Kruuse}, {Kipper}, {Tuvikene},
  {Sorce}  \& {Stoica}}{{Tempel} et~al.}{2018}]{Tempel2018}
{Tempel} E.,  {Kruuse} M.,  {Kipper} R.,  {Tuvikene} T.,  {Sorce} J.~G.,
  {Stoica} R.~S.,  2018, \mn@doi [\aap] {10.1051/0004-6361/201833217}, \href
  {https://ui.adsabs.harvard.edu/abs/2018A&A...618A..81T} {618, A81}

\bibitem[\protect\citeauthoryear{{Tikhonov}, {Gottl{\"o}ber}, {Yepes}  \&
  {Hoffman}}{{Tikhonov} et~al.}{2009}]{Tikhonov2009}
{Tikhonov} A.~V.,  {Gottl{\"o}ber} S.,  {Yepes} G.,   {Hoffman} Y.,  2009,
  \mn@doi [Monthly Notices of the Royal Astronomical Society]
  {10.1111/j.1365-2966.2009.15381.x}, \href
  {https://ui.adsabs.harvard.edu/abs/2009MNRAS.399.1611T} {399, 1611}

\makeatother
\end{thebibliography}
\end{document}